\documentclass[lettersize,journal]{IEEEtran}

\usepackage{amsmath,amsfonts,graphics,amssymb,epsfig,subfigure,color,cite}                      

\usepackage{amsthm,textcomp}
\theoremstyle{plain}
\newtheoremstyle{mystyle}
  {0mm}
  {0mm}
  {}
  {4mm}
  {\bfseries}
  {:}
  { }
  {\thmname{#1}\thmnumber{ #2}\thmnote{ (#3)}}
\theoremstyle{mystyle}
\usepackage{bm}
\usepackage{array}
\usepackage{multirow}
\usepackage{enumerate}
\usepackage{algorithm}
\usepackage{algpseudocode}
\usepackage{adjustbox}
\usepackage{algcompatible}
\algblockdefx{FORP}{ENDFORP}[1]%
  {\textbf{for }#1 \textbf{do in parallel}}%
  {\textbf{end}}
\algblockdefx{FOR}{ENDFOR}[1]%
  {\textbf{for }#1 \textbf{do}}%
  {\textbf{end}}
\algblockdefx{noEnd}{noEndEnd}[1]
    {#1}
    {{\Statex}}
\algblockdefx{BULLET}{EndBULLET}{$\bullet$ }{}
\algnotext{EndBULLET}
\algblockdefx{IF}{ENDIF}[1]%
  {\textbf{if }#1 \textbf{then}}%
  {\textbf{end}}
\algnewcommand\algorithmicprocedure{\textbf{function}}
\algnewcommand\FUNC{\item[\algorithmicprocedure]}%
\algnewcommand\algorithmicendprocedure{\textbf{end function}}
\algnewcommand\ENDFUNC{\item[\algorithmicendprocedure]}%
 \usepackage{setspace}
\let\Algorithm\algorithm
\renewcommand\algorithm[1][]{\Algorithm[#1]\setstretch{1.4}}
\usepackage{float}
\usepackage{makecell}
\usepackage{graphicx}

\usepackage[caption=false,font=normalsize,labelfont=sf,textfont=sf]{subfig}
\usepackage[right=0.625in, left=0.625in, top=0.7in, bottom=1.0in]{geometry}
\usepackage{booktabs}

\usepackage{bbm}

\floatname{algorithm}{Algorithm}

\newcommand{\argmax}{\operatornamewithlimits{argmax}}

\makeatletter
\newcommand{\vast}{\bBigg@{4.5}}
\newcommand{\Vast}{\bBigg@{7.5}}
\makeatother

\allowdisplaybreaks

\begin{document}
    \title{Robust Nonlinear Transform Coding: A Framework for Generalizable Joint Source-Channel Coding}

            \author{Jihun Park, Junyong Shin, Jinsung Park, and Yo-Seb Jeon 
    \thanks{Jihun Park, Junyong Shin, Jinsung Park, and Yo-Seb Jeon are with the Department of Electrical Engineering, POSTECH, Pohang, Gyeongbuk 37673, South Korea (e-mail: jihun.park@postech.ac.kr; sjyong@postech.ac.kr; jinsung@postech.ac.kr; yoseb.jeon@postech.ac.kr).}
    }
    \vspace{-2mm}	
    
    \maketitle

    \begin{abstract} 
    This paper proposes robust nonlinear transform coding (Robust-NTC), a generalizable digital joint source-channel coding (JSCC) framework that couples variational latent modeling with channel-adaptive transmission. Unlike learning-based JSCC methods that implicitly absorb channel variations, Robust-NTC explicitly models element-wise latent distributions via a variational objective with a Gaussian proxy for quantization and channel noise, allowing encoder-decoder to capture latent uncertainty without channel-specific training. Using the learned statistics, Robust-NTC also facilitates rate-distortion optimization to adaptively select element-wise quantizers and bit depths according to online channel conditions.
    To support practical deployment, Robust-NTC is integrated into an orthogonal frequency-division multiplexing (OFDM) system, where a unified resource allocation framework jointly optimizes latent quantization, bit allocation, modulation order, and power allocation to minimize transmission latency while guaranteeing learned distortion targets. \textcolor{black}{Simulation results demonstrate that for practical OFDM systems, Robust-NTC achieves superior rate-distortion efficiency and stable reconstruction fidelity compared to both a conventional separated coding scheme and digital JSCC baselines across various channel conditions.}
    \end{abstract}
    
    \begin{IEEEkeywords}
    Joint source-channel coding, nonlinear transform coding, semantic communication, orthogonal frequency-division multiplexing (OFDM).
    \end{IEEEkeywords}
 
\section{Introduction}\label{Sec:Intro}
Joint source-channel coding (JSCC) has emerged as a powerful paradigm for achieving efficient and adaptive communication under practical constraints on latency, reliability, and coding resources.
By jointly optimizing the source and channel coding processes within a unified rate-distortion framework, JSCC overcomes the fundamental limitations of the classical separation principle, whose theoretical optimality holds only in the asymptotic regime of infinite block lengths~\cite{Waterfilling, Waterfilling_2}.
In practical scenarios, where finite block lengths and computational restrictions limit the effectiveness of channel codes~\cite{Finite_Capacity}, entropy-coding-based source compression~\cite{uniform_proxy, NTC, Hyperprior} becomes highly vulnerable to transmission errors, as even a single bit flip can lead to catastrophic decoding failures.
This fragility, combined with the mismatch between compression rate and channel reliability, results in severe degradation of end-to-end performance in separated coding systems.
Early studies have addressed these issues by formulating analytical rate-distortion optimization problems under idealized source distributions such as Gaussian or Laplacian models~\cite{AnalogJSCC_old, AnalogJSCC_old2, Quant_Design_Lloyd, Quant_Design_opt}.
Analog JSCC schemes~\cite{AnalogJSCC_old, AnalogJSCC_old2} directly map continuous source signals onto channel waveforms, providing robustness without explicit coding but lacking compatibility with digital systems.
Digital JSCC approaches~\cite{Quant_Design_Lloyd, Quant_Design_opt} incorporate quantization for discrete transmission and derive optimal quantizers for noisy channels.
While fundamental, these classical methods are largely theoretical and lack the capability to handle the complexity of high-dimensional, real-world data.

Recent advances in deep learning have enabled learning-based JSCC approaches, which directly learn from realistic data distributions rather than relying on predefined analytical models~\cite{OFDM_RL, semantic_text, DeepJSCC}.
The common idea of the learning-based JSCC approaches is to adopt end-to-end optimization to minimize task-specific distortion under stochastic channel conditions.
Based on this idea, autoencoder-based JSCC architectures for image transmission have been studied in~\cite{DeepJSCC, Analog_mix, NTSCC}, where an image is encoded into a compact latent representation, transmitted through a noisy channel, and reconstructed by a neural decoder.
The work in~\cite{DeepJSCC} introduces the foundational DeepJSCC model which offers a remarkable advantage in transmitting small images over short block lengths, outperforming traditional separated schemes.
Following this, the work in~\cite{Analog_mix} formulates the communication process within a variational autoencoder (VAE) framework, interpreting the channel as a stochastic sampling operation.
This formulation reveals that the VAE loss function naturally upper-bounds the rate-distortion objective, enabling robust JSCC learning that generalizes across varying channel conditions.
Subsequent research~\cite{NTSCC} introduces entropy-based feature-length adaptation, enabling variable-length transmission and achieving high reconstruction quality even for complex, high-resolution content.
Despite these efforts, most existing DeepJSCC models~\cite{DeepJSCC, Analog_mix, NTSCC} depend on gradient-based optimization in a continuous-valued signal space, making them difficult to directly integrate with digital transceivers that rely on discrete constellations and quantized representations.

To bridge this gap, digital DeepJSCC frameworks have been developed by incorporating discrete operations into end-to-end learning pipelines, which are applied directly to the compressed latent representation, including quantization and modulation.
Representative directions include scalar quantization~\cite{sDMCM, Blind, OFDM_ICT}, vector quantization with learned codebooks~\cite{MVQ, sDAC, univasalVQ}, and direct mapping of latent features onto fixed QAM constellations~\cite{MDJCM}.
These approaches improve compatibility with digital transceivers and narrow the gap between learned representations and realizable signaling.
Despite this progress, a key practical bottleneck remains: many designs implicitly overfit to the channel conditions encountered during training, making them sensitive to channel mismatch and often requiring retraining for new environments.
To mitigate this issue, recent studies have introduced channel-adaptive mechanisms, including signal-to-noise (SNR) conditioned networks~\cite{univasalVQ, MDJCM} and adaptive transmission schemes that enforce pretrained bit-error-rate (BER) targets~\cite{Blind, OFDM_ICT, MVQ}.
However, these methods rely on a fixed or limited set of quantization and modulation settings, making them suboptimal for diverse latent distributions across source samples and channels with large dynamic variation.
Furthermore, most prior DeepJSCC frameworks rely on black-box, data-driven training to implicitly learn channel adaptation.
Such models must internalize an enormous number of channel realizations to generalize over varying conditions, leading to excessive model size, high training cost, and limited scalability.
Consequently, achieving efficient and robust channel adaptation through pure end-to-end learning becomes increasingly impractical as channel dynamics and SNR variations grow.

Fundamentally, most existing DeepJSCC frameworks lack an explicit probabilistic model of the latent source distribution that governs the transmitted representation.
Without such modeling, it becomes difficult to analytically characterize the latent-level distortion caused by quantization and channel perturbations, which influences end-to-end performance \cite{MVQ, OFDM_RL} and must be accurately evaluated for effective rate-distortion optimization.
In contrast, classical JSCC studies~\cite{AnalogJSCC_old, AnalogJSCC_old2, Quant_Design_Lloyd, Quant_Design_opt} achieve analytical optimization by assuming simplified yet tractable source distributions.
To realize efficient rate-distortion optimization under realistic and diverse channel conditions, two key requirements arise.
First, the framework must learn a probabilistic latent model that provides an interpretable representation of high-dimensional data, allowing it to approximate a tractable source distribution that captures both statistical variability and distortion sensitivity of individual latent elements.
Such modeling enables transmission parameters to be optimized based on analytical relationships rather than purely empirical fitting.
Second, the inferred latent distributions must be efficiently shared and consistently interpreted between the transmitter and receiver to ensure coherent probabilistic inference and reliable reconstruction under dynamically varying channel conditions.

This paper introduces robust nonlinear transform coding (Robust-NTC), a generalizable digital JSCC framework that achieves rate-distortion-based adaptive transmission by decoupling representation learning from specific channel conditions.
Building on the principle of nonlinear transform coding (NTC)~\cite{NTC}, Robust-NTC formulates the latent representation as a set of probabilistic variables whose parameters explicitly capture element-wise uncertainty and distortion sensitivity.
A Gaussian-proxy formulation is introduced within a variational learning framework, allowing the model to learn latent statistics that reflect quantization- and channel-induced variations without assuming any channel model during training.
To ensure consistent probabilistic inference between the transmitter and receiver, the framework further incorporates a shared hyperprior~\cite{NTC} that transmits compact side information representing the latent mean and variance with negligible overhead.
During transmission, a distortion-matching strategy exploits the learned statistics to determine, for each latent element, the optimal quantizer and bit depth that satisfy its target distortion under the realized channel state.
Finally, the framework is implemented within a standards-compliant orthogonal frequency-division multiplexing (OFDM) transceiver, where a unified optimization procedure jointly allocates quantization bits, modulation order, subcarrier power, and symbol resources.
Through this integration, Robust-NTC achieves scalable and efficient channel adaptation, minimizing OFDM symbol consumption while maintaining distortion-consistent reconstruction quality across diverse fading and SNR conditions.
The main contributions of this paper are summarized as follows:
\begin{itemize}
    \item We propose a generalizable digital JSCC framework that decouples representation learning from specific channel conditions and enables adaptive transmission control through rate-distortion-based optimization.

    \item We develop a variational latent modeling formulation in which both quantization and channel corruption are represented as Gaussian proxy perturbations.  
    This formulation allows the network to learn element-wise latent distributions that encode uncertainty and distortion sensitivity without assuming any specific channel model.  
    A shared hyperprior is incorporated to transmit compact side information, ensuring consistent probabilistic inference between the transmitter and receiver.
    
    \item We introduce a distortion-matching adaptation mechanism that optimizes the quantizer and bit depth of each latent element according to the learned distortion statistics and instantaneous channel conditions.  
    This mechanism enables channel-adaptive rate-distortion control without retraining, achieving efficient and reliable adaptation across diverse SNRs.
    
    \item We extend the proposed framework to a standards-aligned OFDM transceiver by formulating a cross-layer rate-distortion optimization problem that minimizes the number of required OFDM symbols while satisfying latent-level distortion constraints.  
    The proposed solution jointly optimizes latent quantization, bit allocation, subcarrier modulation, and power allocation.  
    To solve this problem efficiently, we design a two-stage algorithm that separately handles source-side distortion-aware bit allocation and channel-side modulation-power adaptation.
    
    \item \textcolor{black}{We provide comprehensive experimental validation on CIFAR-10, STL-10, and Kodak datasets under 3GPP TDL-C fading channels \cite{3GPP_TR_38_901_v18}. 
    The results demonstrate that Robust-NTC achieves superior reconstruction quality and rate efficiency compared to both a conventional separated coding scheme and existing digital DeepJSCC methods, while maintaining strong robustness to frequency-selective fading and stable performance across diverse SNR conditions.}
    
\end{itemize}

\section{System Model and Preliminary Analysis}\label{Sec:System}
This section formalizes the general digital communication system model, details the end-to-end JSCC pipeline, and provides a preliminary analysis that forms the basis of our framework.
\subsection{System Model}\label{Sec:System_model}
We consider a JSCC system operating over a digital communication link. At the transmitter, the source image ${\mathbf{x}}\in\mathbb{R}^{C\times H\times W}$ is processed by a nonlinear analysis transform ${g}_{a}({\mathbf{x}};\boldsymbol{\phi}_g)$, which generates a latent vector ${\mathbf y}=[y_1,\ldots,y_N]^{\sf T}\in\mathbb{R}^N$ representing the semantic features of the input. Since the latent variables are continuous-valued, they must be quantized prior to digital transmission.
Each element $y_i$ is quantized by a scalar quantizer $\mathcal{Q}_i(\cdot)$, which partitions the real line into $L_i$ disjoint intervals: $\mathcal{I}_{i,l} = (T_{i,l-1},\,T_{i,l}],$ for $l \in \{1,\ldots,L_i\}$, 
where $T_{i,0}=-\infty$ and $T_{i,L_i}=+\infty$ denote the boundary values, and 
$\mathcal{T}_i = \{T_{i,1}, \ldots, T_{i,L_i-1}\}$ represents the set of quantization thresholds.
Unlike entropy-coded compression methods such as JPEG or NTC \cite{NTC} that use variable-length codes and are highly sensitive to channel errors, our framework adopts fixed-length codeword mapping as in prior JSCC approaches~\cite{Quant_Design_Lloyd, Quant_Design_opt}.
Let $\mathbf{u}_{i,l}\in\{0,1\}^{b_i}$ be a binary codeword associated with the interval $\mathcal{I}_{i,l}$, such that $L_i \leq 2^{b_i}$. 
Formally, the output of the scalar quantizer is expressed as 
\begin{align}
\mathbf{c}_i \triangleq  \mathcal{Q}_i(y_i)= \sum_{l=1}^{L_i} \mathbf{u}_{i,l} \mathbb{I}(y_i \in \mathcal{I}_{i,l}),
\end{align}
where $\mathbb{I}(\cdot)$ is the indicator function. 
By concatenating $\{\mathbf{c}_i\}_{i=1}^N$, the transmit bitstream is obtained as
\begin{align}
    \mathbf{c} =  \mathrm{Concat}(\mathbf{c}_1,\ldots,\mathbf{c}_N).
\end{align}

In the digital communication framework, the bitstream $\mathbf{c}$ is conveyed through a noisy channel, which we model as a stochastic transfer function. Specifically, the received sequence is given by
\begin{align}
    \hat{\mathbf{c}} = {\mathcal W}(\mathbf{c};\boldsymbol{\eta}),
\end{align}
where ${\mathcal W}(\cdot;\boldsymbol{\eta})$ denotes the channel transfer function parameterized by $\boldsymbol{\eta}$, which captures the effects of system parameters such as noise statistics, fading statistics (e.g., Rayleigh fading or multipath propagation in OFDM), transmit power, modulation, and coding schemes.
The transition probability associated with the channel transfer function is expressed as
\begin{align}
    p_{\hat{\mathbf{c}}|\mathbf{c}}(\hat{\mathbf{c}}|\mathbf{c})
    = \Pr\!\left[{\mathcal W}(\mathbf{c};\boldsymbol{\eta})=\hat{\mathbf{c}}\right].
\end{align}
At the receiver, the corrupted codeword $\hat{\mathbf{c}}_i$ is mapped to a reconstruction level $R_{i,q}$ from the codebook ${\mathcal R}_i=\{R_{i,1},R_{i,2},\ldots,R_{i,2^{b_i}}\}$: 
\begin{align}
    \hat{y}_i \triangleq \mathcal{Q}_i^{-1}(\hat{\mathbf{c}}_i) = \sum_{q=1}^{2^{b_i}} R_{i,q} \mathbb{I}(\hat{\mathbf{c}}_i = \hat{\mathbf{u}}_{i,q}),
\end{align}
where $\hat{\mathbf{u}}_{i,q}\in\{0,1\}^{b_i}$ denotes the received binary codeword associated with the reconstruction level $R_{i,q}$.
The latent estimate $\hat{\bf y}=[\hat{y}_1,\ldots,\hat{y}_N]^{\sf T}$ is converted back to the image domain by the nonlinear synthesis transform  ${g}_{s}(\hat{\bf y};\boldsymbol{\theta}_g)$, yielding the reconstructed image
\begin{align}
    \hat{\mathbf{x}} = g_s(\hat{\bf y};\boldsymbol{\theta}_g) \in \mathbb{R}^{C\times H\times W}.
\end{align}
The overall end-to-end pipeline depends not only on the quantizer design for each latent element, but also on the choice of channel parameters $\boldsymbol{\eta}$. These aspects, including optimal quantizer construction and channel-aware parameter optimization, will be discussed in the sequel.

\subsection{Distortion Analysis}\label{Sec:Analysis}
In general, the end-to-end performance of DeepJSCC is primarily influenced by the distortion of the compressed latent representation \cite{MVQ, OFDM_RL}.
Motivated by this observation, we initiate our analysis by characterizing the expected mean-squared error (MSE) of individual latent elements as:
\begin{align}
    D_i \triangleq \mathbb{E}\!\left[(y_i - \hat{y}_i)^2\right], 
    \qquad i\in\{1,\ldots,N\}.
\end{align}
In general, analyzing the distortion induced by the channel transfer function ${\mathcal W}(\cdot;\boldsymbol{\eta})$ is challenging because the reconstruction error depends on the joint distribution of bit errors across an entire codeword. To obtain a tractable model, we approximate the channel transfer function by a bank of parallel binary symmetric channels (BSCs). Accordingly, the transition probability for the codeword $\mathbf{c}_i$ is given by
\begin{align}
    p_{\hat{\mathbf{c}}_i|\mathbf{c}_i}(\hat{\mathbf{c}}_i|\mathbf{c}_i)
    &= \Pr\!\left[{\mathcal W}(\mathbf{c}_i;\boldsymbol{\eta})=\hat{\mathbf{c}}_i\right] \nonumber \\
    &\approx \prod_{j=1}^{b_i}
    \epsilon_{i,j}^{\,\mathbb{I}(c_{i,j}\neq \hat{c}_{i,j})}
    (1-\epsilon_{i,j})^{\,\mathbb{I}(c_{i,j}=\hat{c}_{i,j})},
    \label{eq:BSC_transition_ci}
\end{align}
where $\epsilon_{i,j} \in [0,1/2]$ denotes the bit-flip probability of the $j$-th bit in $\mathbf{c}_i$.
Under the above model, the transmission chain for a latent element can be described as
\begin{align}{
y_i \;\xrightarrow[\text{quantization}]{\;\mathcal{Q}_i\;}\; \mathbf{c}_i 
\;\xrightarrow[\text{channel errors}]{\;\text{BSC w/ }\boldsymbol{\epsilon}_i\;}\; \hat{\mathbf{c}}_i
\;\xrightarrow[\text{dequantization}]{\;\mathcal{Q}_i^{-1}\;}\; \hat{y}_i},\label{eq:transmission_chain}
\end{align}
where $\boldsymbol{\epsilon}_i=[\epsilon_{i,1},\ldots,\epsilon_{i,b_i}]^{\sf T}$ collects the bit-flip probabilities for the $b_i$ bits of $\mathbf{c}_i$. 
\textcolor{black}{The expected distortion of the latent element $y_i$ is then systematically derived by evaluating the continuous expectation over disjoint quantization intervals. For a given source value $y \in \mathcal{I}_{i,l}$, the encoded codeword is deterministically fixed as $\mathbf{c}_i = \mathbf{u}_{i,l}$. Thus, the reconstructed value $\hat{y}_i = R_{i,q}$ is solely governed by the channel transition probability from $\mathbf{u}_{i,l}$ to $\hat{\mathbf{u}}_{i,q}$. Accordingly, the overall expected distortion is expanded as follows:}
\begin{align}
    \textcolor{black}{D_i} 
    &\textcolor{black}{= \mathbb{E}\big[ (y_i - \hat{y}_i)^2 \big]} \nonumber \\
    &\textcolor{black}{= \sum_{l=1}^{L_i} \int_{\mathcal{I}_{i,l}} \mathbb{E}\big[ (y - \hat{y}_i)^2 \mid \mathbf{c}_i = \mathbf{u}_{i,l} \big] p_{y_i}(y) \, dy} \nonumber \\
    &\textcolor{black}{= \sum_{l=1}^{L_i} \int_{\mathcal{I}_{i,l}} \left( \sum_{q=1}^{2^{b_i}} (y - R_{i,q})^2 P\!\left(q\,\big|\,l;\,\boldsymbol{\epsilon}_i\right) \right) p_{y_i}(y) \, dy} \nonumber \\
    &\textcolor{black}{= \sum_{l=1}^{L_i} \sum_{q=1}^{2^{b_i}} P\!\left(q\,\big|\,l;\,\boldsymbol{\epsilon}_i\right) \int_{\mathcal{I}_{i,l}} \big(y - R_{i,q}\big)^2 p_{y_i}(y)\,dy,}
    \label{eq:Di_mse}
\end{align}
where $p_{y_i}(y)$ is the probability density function (PDF) of $y_i$, and 
\begin{align}
    P\!\left(q\,\big|\,l;\,\boldsymbol{\epsilon}_i\right)
    &\approx
    \prod_{j=1}^{b_i}
    \epsilon_{i,j}^{\,\mathbb{I}(u_{i,l}^{(j)} \neq u_{i,q}^{(j)})}\,
    (1-\epsilon_{i,j})^{\,\mathbb{I}(u_{i,l}^{(j)} = u_{i,q}^{(j)})},
    \label{eq:Pi_product}
\end{align}
denotes the approximated probability that the codeword $\mathbf{u}_{i,l}$ corresponding to region $\mathcal{I}_{i,l}$ is decoded as $\mathbf{u}_{i,q}$, with $u_{i,l}^{(j)}$ the $j$-th bit of $\mathbf{u}_{i,l}$.

\section{Robust Nonlinear Transform Coding Framework for Generalizable JSCC}\label{Sec:RobustNTC}
This section presents the Robust-NTC framework, integrating channel-optimized quantizers, variational modeling, and distortion-matching with element-wise targets for robust end-to-end communication.

\subsection{Motivation}
The distortion analysis in \eqref{eq:Di_mse} reveals that the per-latent distortion is governed by three factors: (i) the latent source distribution $p_{y_i}$, (ii) the channel-induced flip probabilities $\boldsymbol{\epsilon}_i$, and (iii) the quantizer design, including the bit allocation $b_i$, thresholds, reconstruction levels, and codeword mapping. 
These dependencies expose a limitation of prior digital DeepJSCC designs: a fixed quantizer~\cite{OFDM_RL, sDMCM, Blind} or a learnable codebook optimized for a fixed channel condition~\cite{MVQ, sDAC, univasalVQ} is not optimal under varying channel conditions, and may result in excessive distortion under adverse conditions.

Based on the above observations, we put forward the following design questions:
\begin{itemize}
    \item How can quantizers and dequantizers be adaptively optimized for each channel realization while jointly accounting for heterogeneous feature distributions across sources and among latent elements within a source?
    
    \item How can more informative latent features be prioritized in bit allocation and channel reliability (i.e., through the joint optimization of $b_i$ and $\boldsymbol{\epsilon}_i$) to enable efficient end-to-end communication?
    More specifically, what distortion level should be guaranteed for each latent element?  

\end{itemize}

To address the above questions, we propose a Robust-NTC framework based on a joint modeling-and-control paradigm, which (i) learns distortion-aware latent statistics through variational learning (i.e., training period) and (ii) enforces element-wise distortion targets during transmission (i.e., inference period).
The proposed framework enables end-to-end training without dependence on specific channel or codebook configurations, thereby facilitating adaptive quantizer design and distortion-consistent resource allocation across diverse channel conditions.

\subsection{Optimal Quantizer Design for Noisy Channels via Gaussian Modeling}\label{Sec:Quantizer}
Formally, the optimal quantizer design problem for a given bit allocation $b_i$ and flip-probability vector $\boldsymbol{\epsilon}_i$ is expressed as\footnote{The optimal quantizer design should jointly optimize both the quantization rule and the codeword mapping. However, since the search space of mappings is combinatorial and computationally intractable, we simplify the problem by optimizing thresholds and reconstruction levels from multiple random initializations and retaining the best converged solution.}
\begin{align}
\min_{{\mathcal T}_i,\,{\mathcal R}_i} 
\sum_{\ell=1}^{L_i} \sum_{q=1}^{2^{b_i}}
P\left(q\,\big|\,\ell;\,\boldsymbol{\epsilon}_i\right)\int_{\mathcal{I}_{i,\ell}}
\big(y - R_{i,q}\big)^2p_{y_i}(y)\,dy.
\label{eq:Opt_yi_refined}
\end{align} 
The formulation in \eqref{eq:Opt_yi_refined} shows that the optimal thresholds and reconstruction levels depend on the latent source distribution $p_{y_i}$.
Without an explicit and tractable model for $p_{y_i}$, the integral
terms in \eqref{eq:Opt_yi_refined} cannot be evaluated in closed form, making
the optimization intractable—particularly when the design must be adapted across
heterogeneous latent elements and varying channel conditions.

\begin{figure}[t]
    \centering
    \begin{minipage}{\linewidth}
        \centering
        \subfigure[\textcolor{black}{Optimal quantizer for the normalized source $\bar{y} \sim \mathcal{N}(0,1)$}]{
            \includegraphics[width=7.2cm]{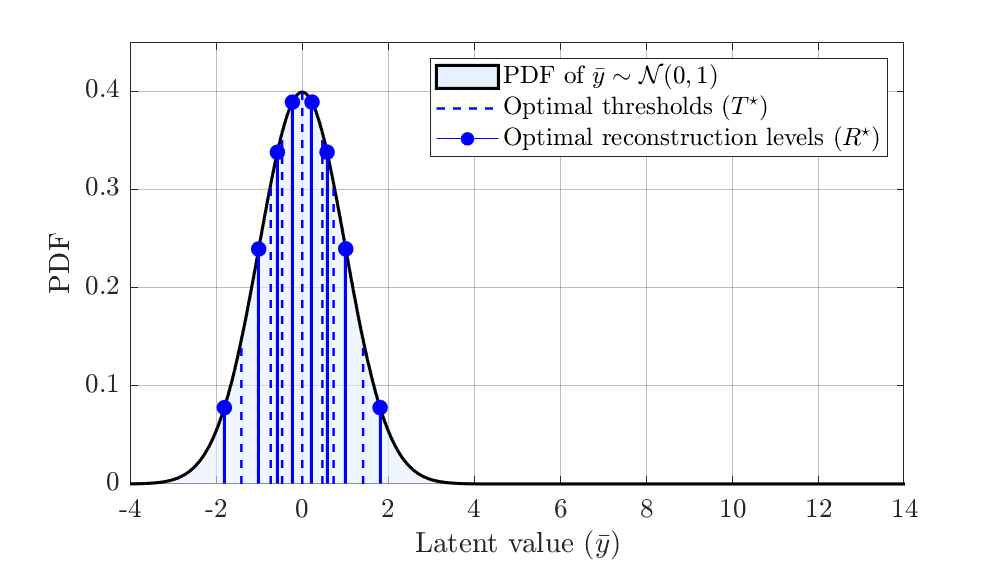}
        }
    \end{minipage}

    \vspace{-0mm} 

    \begin{minipage}{\linewidth}
        \centering
        \subfigure[\textcolor{black}{Transformed quantizer for the target source $y_i \sim \mathcal{N}(\mu_i, \sigma_i^2)$}]{
            \includegraphics[width=7.2cm]{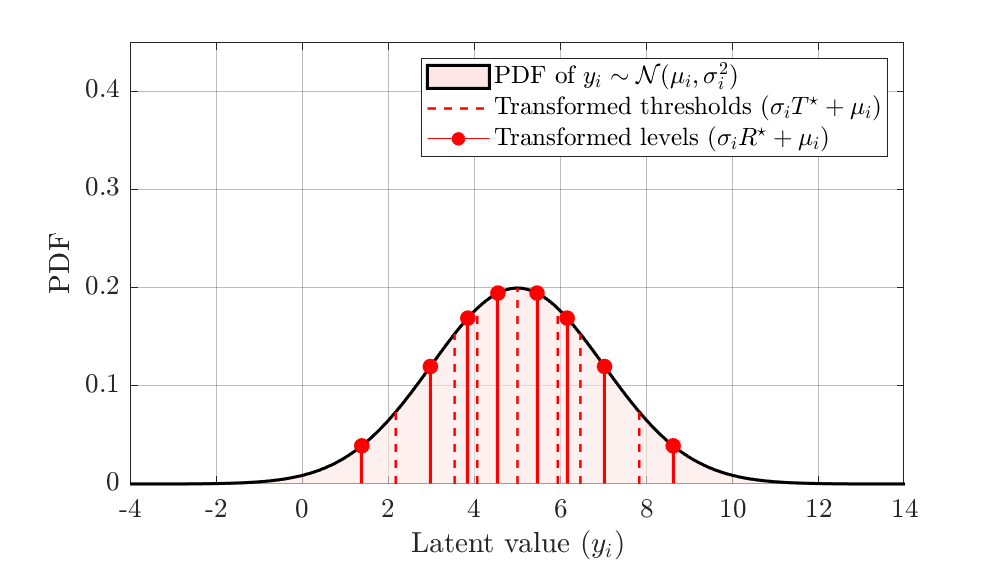}
        }
    \end{minipage}
    \caption{\textcolor{black}{An illustrative example of the affine transformation rule for quantizer design. (a) Thresholds and reconstruction levels designed for the normalized Gaussian source $\bar{y}$, with $b=3$ and $\boldsymbol{\epsilon} = [0.01, 0.01, 0.01]^{\sf T}$. (b) Transformed quantizer parameters for the target source $y_i$ with $\mu_i=5$ and $\sigma_i=2$.}}
    \label{fig:Quantizer_Transformation}
\end{figure}

To enable a tractable yet effective design, we introduce a Gaussian modeling assumption for each latent element, namely $y_i \sim \mathcal{N}(\mu_i,\sigma_i^2)$. 
The learning of the Gaussian latent structure and the transmission of $(\mu_i,\sigma_i^2)$ will be detailed in Sec.~\ref{Sec:Variational_Modeling}.
Under the Gaussian assumption, a key observation is that, for identical $\boldsymbol{\epsilon}_i$ and $b_i$, the optimal thresholds and reconstruction levels for $y_i$ are affine transforms of those for a normalized source $\bar{y}\sim\mathcal{N}(0,1)$. \textcolor{black}{Specifically, as illustrated in Fig.~\ref{fig:Quantizer_Transformation}, if $\{T^\star_\ell\}_{\ell=1}^{L_i-1}$ and $\{R^\star_q\}_{q=1}^{2^{b_i}}$ denote the optimal thresholds and reconstruction levels for ${\bar y}$, then the corresponding parameters for $y_i$ are $\{\sigma_i T^\star_\ell+\mu_i\}_{\ell=1}^{L_i-1}$ and $\{\sigma_i R^\star_q+\mu_i\}_{q=1}^{2^{b_i}}$, respectively.}
Therefore, without loss of generality, we can remove the index $i$ from \eqref{eq:Opt_yi_refined} and rewrite the optimization problem for the normalized Gaussian source $\bar y$ as 
\begin{align}
\min_{{\mathcal T},\,{\mathcal R}} 
\sum_{\ell=1}^{L} \sum_{q=1}^{2^{b}}
P\left(q\,\big|\,\ell;\,\boldsymbol{\epsilon}\right)
\int_{\mathcal{I}_{\ell}}\,
\big(\bar y - R_{q}\big)^{2}
p_{\bar y}(\bar y)\,d\bar y,
\label{eq:normalized_problem}
\end{align}
where $b$ is the bit depth and $\boldsymbol{\epsilon}$ denotes the channel condition.
The design of scalar quantizers for Gaussian sources over BSCs has been extensively investigated~\cite{Quant_Design_Lloyd, Quant_Design_opt}. 
In particular, the approach in~\cite{Quant_Design_Lloyd} extends the classical Lloyd-Max algorithm by incorporating channel transition probabilities into the centroid and boundary updates.
However, unlike the noiseless case, this iterative process does not guarantee monotonic distortion reduction.
To address this,~\cite{Quant_Design_opt} derives sufficient optimality conditions and proposes an iterative procedure that converges to a locally optimal solution.
We adopt the methodology in~\cite{Quant_Design_opt} and summarize the two alternating steps below.
\begin{itemize}
\item {\bf Thresholds update:}
For fixed reconstruction levels ${\mathcal R}$, the optimal quantization regions $\mathcal{I}_l^{\star}$ partition the source domain to minimize the expected end-to-end distortion when mapping ${\bar y}$ to codeword $\mathbf{u}_l$:
\begin{align}
    \mathcal{I}_l^{\star} &= \Big\{{\bar y}: {\mathbb E}[({\bar y}-\hat{y})^2\mid \mathbf{u}_l] \le {\mathbb E}[({\bar y}-\hat{y})^2\mid \mathbf{u}_j], \forall j \neq l \nonumber\Big\} \\ &=\Big\{{\bar y}: 2{\bar y}\Big({\mathbb E}[\hat{y}\mid \mathbf{u}_j] - {\mathbb E}[\hat{y}\mid \mathbf{u}_l]\Big) \nonumber\\ &\qquad \qquad \qquad \qquad \le {\mathbb E}[\hat{y}^2\mid \mathbf{u}_j] - {\mathbb E}[\hat{y}^2\mid \mathbf{u}_l], \forall j \neq l \Big\}.\label{eq:Reconstruction_update}
\end{align}
The optimal boundary $T_l^{\star}$ between two adjacent active regions, $\mathcal{I}_l^{\star}$ and $\mathcal{I}_{l+1}^{\star}$, is the point where the expected distortions are equal. 
This process identifies and discards ``useless'' codewords for which the corresponding partition region is empty, an effect that becomes more pronounced as the channel degrades.\footnote{This motivates the setting $L \leq 2^{b}$ in this work, since some codewords may become inactive under severe channel degradation.}

\item {\bf Reconstruction levels update:}
For fixed thresholds ${\mathcal T}$, the optimal reconstruction level for a received codeword $\hat{\mathbf{u}}_q$ is given by the minimum mean square error (MMSE) estimate of $\bar y$ conditioned on $\hat{\mathbf{u}}_q$:
\textcolor{black}{
    \begin{align}
        {R_q^{\star}}
        &{= \mathbb{E}[\bar y \mid \hat{\mathbf{u}}_q]} \nonumber \\
        &{= \frac{\sum_{\ell=1}^{L} P\!\left(q \,\big|\, \ell;\boldsymbol{\epsilon}\right)\int_{T_{\ell-1}}^{T_\ell} \bar y\, \phi(\bar y)\, d\bar y}
               {\sum_{\ell=1}^{L} P\!\left(q \,\big|\, \ell;\boldsymbol{\epsilon}\right)\int_{T_{\ell-1}}^{T_\ell} \phi(\bar y)\, d\bar y}} \nonumber \\
        &\overset{(a)}{=} \frac{\sum_{\ell=1}^L P\!\left(q \,\big|\, \ell;\boldsymbol{\epsilon}\right)\,
              \big(\phi(T_{\ell-1})-\phi(T_\ell)\big)}
               {\sum_{\ell=1}^L P\!\left(q \,\big|\, \ell;\boldsymbol{\epsilon}\right)\,
              \big(\Phi(T_\ell)-\Phi(T_{\ell-1})\big)},
        \label{eq:Rq_update_closed}
    \end{align}
where $\phi(\cdot)$ and $\Phi(\cdot)$ denote the PDF and cumulative distribution function (CDF) of the standard Gaussian distribution, respectively, and $(a)$ follows from its integration property $\int \bar y \phi(\bar y) d\bar y = -\phi(\bar y)$.}
\end{itemize}

\begin{figure}[t]
    \centering
    \includegraphics[width=0.8\columnwidth]{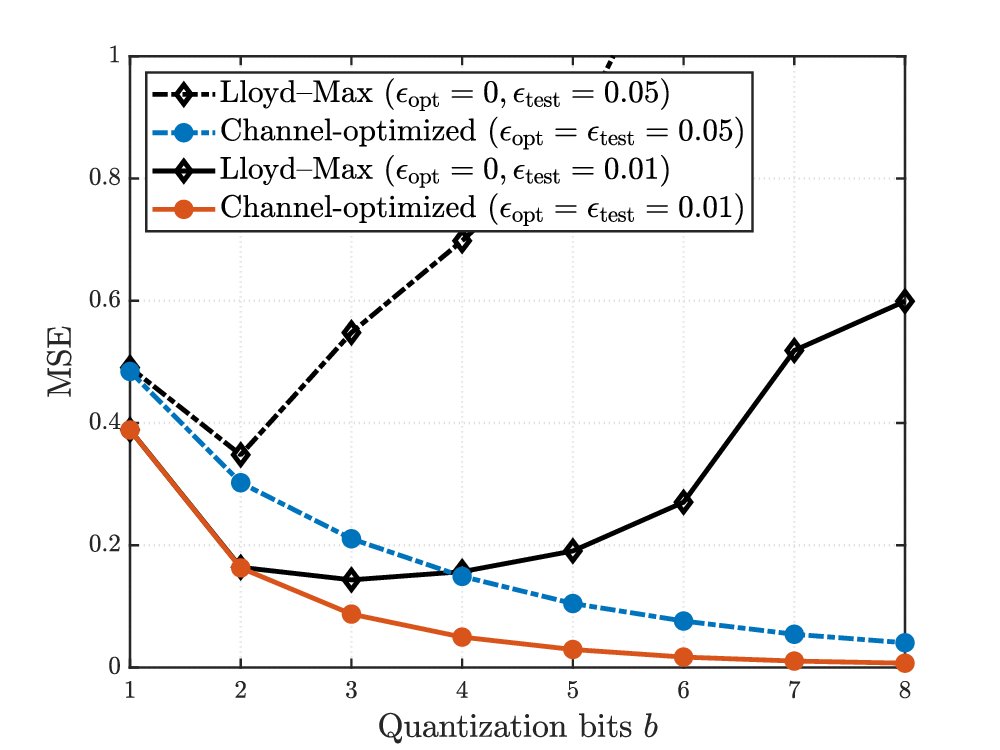}
    \vspace{-3mm}
    \caption{\textcolor{black}{MSE performance of quantizer designs across different bit depths for various optimized-test flip probability pairs $(\epsilon_{\rm opt},\epsilon_{\rm test})$. Specifically, each quantizer is optimized for the target condition $(b,\boldsymbol{\epsilon}=\epsilon_{\rm opt}{\bf 1})$, and the resulting distortion is evaluated under the test condition $(b,\boldsymbol{\epsilon}=\epsilon_{\rm test}{\bf 1})$, where ${\bf 1}\in\mathbb{R}^b$ denotes the all-ones column vector.}}
    \label{fig:quantizer_opt_merge}
\end{figure}
\textcolor{black}{By iterating between these two update rules until convergence, we obtain a locally optimal channel-optimized quantizer for the target condition $(b,\boldsymbol{\epsilon})$.}
For each Gaussian latent $y_i \sim \mathcal{N}(\mu_i,\sigma^2_i)$, 
the resulting design yields a distortion that scales linearly with the source variance:
\begin{align}
D^{\star}(\sigma^2_i; b,\boldsymbol{\epsilon})
= \sigma^2_i D^{\star}(1; b,\boldsymbol{\epsilon}),
\end{align}
where $D^{\star}(1; b,\boldsymbol{\epsilon})$ denotes the MSE of the normalized Gaussian source under the quantizer optimized for $(b,\boldsymbol{\epsilon})$.
Fig.~\ref{fig:quantizer_opt_merge} compares Lloyd–Max quantizers with 
channel-optimized quantizers across different bit depths.
Across all tested channel conditions, the channel-optimized quantizers consistently achieve lower MSE than Lloyd-Max quantizers, whose distortion increases sharply with bit depth under channel noise. 
Moreover, for channel-optimized quantizers, the results show that higher flip probabilities require more quantization bits to attain the same distortion level, highlighting the need for channel-adaptive allocation.

These observations imply that latent elements with larger variances $\sigma_i^2$ must be assigned more bits to achieve the same distortion level, consistent with classical rate-distortion theory. 
At the same time, noisier channels demand additional resources to compensate for the increased probability of errors. 
Although $D^{\star}(\sigma^2; b,\boldsymbol{\epsilon}) \to 0$ as $b \to \infty$, the required resources grow excessively. Hence, it is essential to define element-wise distortion targets that balance accuracy and efficiency under both source statistics and channel conditions.
The following subsections introduce a variational framework that jointly learns the latent distribution and its distortion targets, together with matching schemes that ensure consistent alignment between training and inference.

\subsection{Variational Learning of the Robust-NTC Framework}
\label{Sec:Variational_Modeling}
The proposed framework interprets the JSCC process within the paradigm of VAEs~\cite{VAE}, by treating quantization \cite{NTC} and channel noise \cite{Analog_mix} as stochastic sampling applied to the latent variables.
Within this framework, the probabilistic generative and inference models are jointly optimized by minimizing the expected Kullback-Leibler (KL) divergence between the variational posterior and the true posterior:
\begin{align}
    \min_{\boldsymbol{\phi}_g, \boldsymbol{\theta}_g}
    \ \mathbb{E}_{\mathbf{x}\sim p(\mathbf{x})}
    \Big[
        D_{\mathrm{KL}}\!\left(q(\hat{\mathbf{y}}|\mathbf{x})
        \,\big\|\, p(\hat{\mathbf{y}}|\mathbf{x})\right)
    \Big],
    \label{eq:KL_general}
\end{align}
where $\mathbf{x}\sim p(\mathbf{x})$ denotes the source distribution, $q(\hat{\mathbf{y}}|\mathbf{x})$ is the variational posterior, and $p(\hat{\mathbf{y}}|\mathbf{x})$ is the generative likelihood.

To enable element-wise latent modeling and to exploit spatial dependencies among ${\bf y}$, 
we enrich the latent representation with a hyperprior variable ${\bf z}$, following the NTC methodology~\cite{Hyperprior, NTC}. 
Within this structure, each latent element $y_i$ is modeled as a Gaussian random variable 
whose mean–variance parameters are inferred from the hyperprior.
A hyper-analysis transform ${\bf z} = h_a({\bf y};\boldsymbol{\phi}_h)$ extracts side information from the latent space, which is subsequently quantized by a uniform quantizer and entropy-coded.
Owing to its negligible rate contribution, we follow the standard NTC assumption that the quantized latent ${\hat{\bf z}}$ can be transmitted error-free through channel coding and recovered via entropy decoding \cite{NTSCC, MDJCM}.
To preserve differentiability during training, the uniform quantizer is replaced by the uniform noise proxy~\cite{uniform_proxy}:
\begin{align}
    \tilde{\mathbf{z}} = \mathbf{z} + \mathbf{n}_{\rm uni}, 
    \qquad \mathbf{n}_{\rm uni} \overset{\text{i.i.d.}}{\sim}~\mathcal{U}\!\left(-\tfrac{1}{2}, \tfrac{1}{2}\right),
    \label{eq:Uniform_proxy}
\end{align}
while at inference the deterministic quantizer $\hat{\bf z} = \mathcal{Q}_{\mathrm{uni}}({\bf z})$ is applied consistently with entropy decoding.
The hyper-synthesis transform $h_s(\cdot;\boldsymbol{\theta}_h)$ then estimates the element-wise Gaussian parameters $(\boldsymbol{\mu},\boldsymbol{\sigma}^2)$ for the latent variables:
\begin{align}
    (\boldsymbol{\mu}, \boldsymbol{\sigma}^2) 
    =
    \begin{cases}
        h_s(\tilde{\bf z};\boldsymbol{\theta}_h), & \text{training (proxy)} \\[6pt]
        h_s(\hat{\bf z};\boldsymbol{\theta}_h),   & \text{inference (uniform quantizer)}.
    \end{cases}
\end{align}
With the hyperprior in place, the KL objective extends to
\begin{align}
    \min_{\boldsymbol{\phi}_g, \boldsymbol{\phi}_h, \boldsymbol{\theta}_g,  \boldsymbol{\theta}_h}
    \ \mathbb{E}_{\mathbf{x}\sim p(\mathbf{x})}
    \Big[
        D_{\mathrm{KL}}\!\left(q(\hat{\mathbf{y}},\tilde{\mathbf{z}}|\mathbf{x})
        \,\big\|\, p(\hat{\mathbf{y}},\tilde{\mathbf{z}}|\mathbf{x})\right)
    \Big],
    \label{eq:KL_with_hyperprior}
\end{align}
where the joint modeling of $(\hat{\mathbf{y}},\tilde{\mathbf{z}})$ yields adaptive priors for the latent distribution. 
\textcolor{black}{This expression follows directly by applying Bayes' rule, and \eqref{eq:KL_with_hyperprior} can be expanded as}
\begin{align}
    &D_{\mathrm{KL}} \!\left(q(\hat{\mathbf{y}},\tilde{\mathbf{z}}|\mathbf{x})
        \,\big\|\, p(\hat{\mathbf{y}},\tilde{\mathbf{z}}|\mathbf{x})\right) \nonumber \\
    &\!= \mathbb{E}_{q(\hat{\mathbf{y}},\tilde{\mathbf{z}}|\mathbf{x})} \big[\!
        -\log p(\mathbf{x}|\hat{\mathbf{y}},\tilde{\mathbf{z}})
        -\log p(\hat{\mathbf{y}}|\tilde{\mathbf{z}})
        -\log p(\tilde{\mathbf{z}})
    \big] + \text{const.},
    \label{eq:KL_expand}
\end{align}
where the first term is modeled by the squared reconstruction error under the assumption of $\mathcal{N}(\mathbf{x}\mid \hat{\mathbf{x}}, (2\rho_g)^{-1}\mathbf{I})$, \textcolor{black}{with $\rho_g$ denoting the weighting parameter for the reconstruction loss,} while the second and third terms represent the latent rate and the hyperprior rate, respectively.
In particular, the hyperprior rate is defined as the convolution of the prior over the hyper-latent $z_j$ with a uniform distribution \cite{uniform_proxy}:
\begin{align}
    \! p_{\tilde z_j|{\boldsymbol \psi^{(j)}}}(\tilde z_j|{\boldsymbol \psi^{(j)}})
    = \left(p_{z_j|{\boldsymbol \psi^{(j)}}}({\boldsymbol \psi^{(j)}}) 
      *\mathcal{U}\!\left(-\tfrac{1}{2}, \tfrac{1}{2}\right)\right)(\tilde z_j),
    \label{eq:hyperprior_proxy}
\end{align}
where ${\boldsymbol \psi^{(j)}}$ denotes the parameters of the non-parametric density model 
$p_{z_j|{\boldsymbol \psi^{(j)}}}$ that enables the entropy model for $z_j$, 
and $*$ denotes the convolution operator.

\begin{figure}[t]
    \centering
    \begin{minipage}{\linewidth}
        \centering
        \subfigure[System model (inference)]{
            \includegraphics[width=8.8cm]{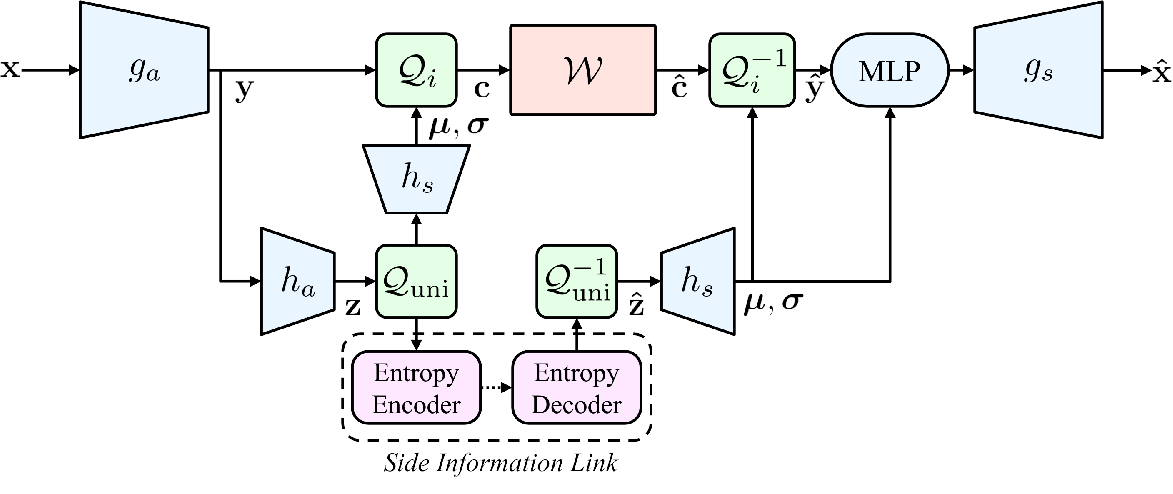}
        }

    \end{minipage}

    \vspace{-0mm} 

    \begin{minipage}{\linewidth}
        \centering
        \subfigure[Proxy model (training)]{
            \includegraphics[width=8.8cm]{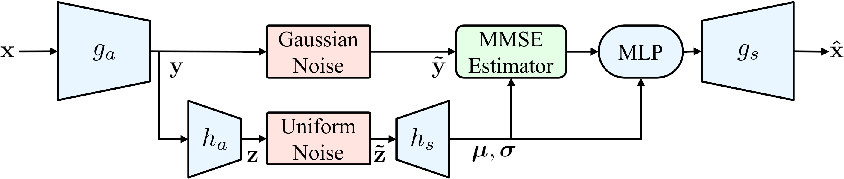}
        }
    \end{minipage}
    \caption{System architecture of the proposed Robust-NTC: (a) overall system model and (b) proxy model for tractable optimization.}
    \label{fig:System_model}
\end{figure}

\textcolor{black}{Up to \eqref{eq:KL_general}--\eqref{eq:hyperprior_proxy}, the formulation follows the standard NTC framework with hyperprior-based variational modeling~\cite{Hyperprior,NTC}.}
In contrast to NTC, which models only the quantization distortion via a uniform-noise proxy under the assumption that the quantized latents of ${\bf y}$ are reliably transmitted~\cite{NTC}, the proposed Robust-NTC approximates the non-differentiable transmission chain of ${\bf y}$, encompassing quantization, channel errors, and dequantization.
\textcolor{black}{This formulation also differs from the prior JSCC framework in~\cite{NTSCC}. Specifically,~\cite{NTSCC} retains the standard NTC variational formulation and applies a JSCC module to the latent representation with entropy-guided variable-length transmission. Whereas, the proposed Robust-NTC explicitly incorporates the effect of channel corruption into the latent modeling framework so as to enable tractable distortion characterization under practical transmission impairments.}
To this end, we introduce a Gaussian proxy for the corrupted latent, which enables differentiable variational training:
\begin{align}
    \tilde{\mathbf{y}} = \mathbf{y} + \mathbf{n}_{G}, 
    \qquad \mathbf{n}_{G}\sim \mathcal{N}(\mathbf{0}, \sigma_d^2 \mathbf{I}),
    \label{eq:Gaussian_proxy}
\end{align}
where $\sigma_d^2$ is a nominal variance parameter. 
\textcolor{black}{The Gaussian proxy is adopted as a tractable and robust surrogate during training. Since the Gaussian distribution attains the maximum differential entropy for a given variance $\sigma_d^2$~\cite{Waterfilling}, it serves as a conservative additive noise model in the variational optimization.}

Without loss of generality, we hereafter set $\sigma_d^2=1$, since any scaling can be absorbed into the latent variance $\sigma_i^2$. 
Given the stochastic parameters $(\mu_i, \sigma_i^2)$ from the hyper-synthesis transform, the noisy latent can be further refined by the MMSE estimator \textcolor{black}{\cite{MMSE_book}}:
\begin{align}
    \tilde y_{\mathrm{MMSE}, i} 
    &= \mathbb{E}[y_i \mid \tilde y_i] \nonumber \\
    &= \mu_i + \frac{\sigma_i^2}{\sigma_i^2+1}\big(\tilde y_i - \mu_i\big).
    \label{eq:mmse_filter}
\end{align}
This estimator removes the predictable component conditioned on $(\mu_i,\sigma_i^2)$ and yields a posterior distribution of the form \textcolor{black}{\cite{MMSE_book}}
\begin{align}
p(y_i \mid \tilde y_i) = \mathcal{N}\left(\tilde y_{\mathrm{MMSE}, i},\frac{\sigma_i^2}{\sigma_i^2+1}\right).
\label{eq:posterior}
\end{align}
Equivalently, the latent can be decomposed as
\begin{align}
y_i
= \tilde y_{\mathrm{MMSE}, i} + \tilde n_{G,i},
\quad \tilde n_{G,i}\sim \mathcal{N}\left(0,\frac{\sigma_i^2}{\sigma_i^2+1}\right),
\label{eq:mmse_equiv}
\end{align}
where the residual variance represents our distortion proxy. 
This variance effectively reflects the combined impact of quantization and channel-induced flips in a quantizer-agnostic manner, and will be used as the target distortion during inference.

Based on the Gaussian proxy in \eqref{eq:Gaussian_proxy}, the conditional prior of the noisy latent $p(\tilde{y}_i|\tilde{\mathbf{z}})$ can be expressed as the convolution of the latent Gaussian distribution with the proxy noise:
\begin{align}
    p_{\tilde{y}_i|\tilde{\mathbf{z}}}(\tilde{y}_i|\tilde{\mathbf{z}}) 
    &= \left(\mathcal{N}(\mu_i, \sigma_i^2) 
       * \mathcal{N}(0,1) \right) (\tilde{y}_i)\nonumber \\
    &= \left(\mathcal{N}(\mu_i, \sigma_i^2+1)\right) (\tilde{y}_i).
    \label{eq:Gaussian_convolution}
\end{align}
The corresponding latent-rate term is given by \textcolor{black}{\cite{MMSE_book}}
\begin{align}
-\log p(\tilde{y}_i | \tilde{\mathbf{z}})
= \frac{1}{2}\log\!\big(2\pi(\sigma_i^2+1)\big)
  + \frac{\big(\tilde{y}_i-\mu_i\big)^2}{2(\sigma_i^2+1)}.
\label{eq:latent_rate_nll} 
\end{align}
The first term penalizes the predictive variance $\sigma_i^2$, and the second term not only enforces consistency between $\tilde{y}_i$ and its mean $\mu_i$ but also regularizes the deviation relative to the scale of $\sigma_i^2$, thereby encouraging the latent distribution to adhere more closely to a Gaussian form. 
By integrating the preceding derivations, the Robust-NTC framework is optimized by minimizing the following rate-distortion Lagrangian:
\begin{align}
    &\mathcal{L}_{\mathrm{RNTC}} \nonumber \\ 
    &=
    \mathbb{E}_{\mathbf{x}\sim p(\mathbf{x})}
    \left[
         \left(
        - \lambda_y \log p_{\tilde{\mathbf{y}}|\tilde{\mathbf{z}}}(\tilde{\mathbf{y}}|\tilde{\mathbf{z}}) -\lambda_z\log p_{\tilde{\mathbf{z}}}(\tilde{\mathbf{z}}) \right)
        + d(\mathbf{x}, \hat{\mathbf{x}})
    \right],
    \label{eq:RNTC_loss}
\end{align}
where $\lambda_y$ controls the rate allocated to the latent representation, $\lambda_z$ regulates the rate dedicated to side information, and $d(\mathbf{x}, \hat{\mathbf{x}})$ denotes the distortion function.
Here, the reconstruction $\hat{\mathbf{x}}$ is obtained as
\begin{align}
\hat{\mathbf{x}}=g_s({\bf \tilde y}_{\mathrm{MMSE}}, \boldsymbol{\mu}, \boldsymbol{\sigma}^2 ;\boldsymbol{\theta}_g),
\end{align}
where the hyperprior statistics $(\boldsymbol{\mu},\boldsymbol{\sigma}^2)$ are first fused with the refined latent $\tilde{\mathbf{y}}_{\rm{MMSE}}$ through a lightweight multi-layer perceptron (MLP) before being passed to the synthesis transform $g_s(\cdot;\boldsymbol{\theta}_g)$.
The overall system architecture and the proxy-based relaxation adopted for tractable optimization are summarized in Fig.~\ref{fig:System_model}.

\subsection{Distortion Matching-Based Adaptive Transmission}\label{Sec:Adaptive_Inference}
During inference, the proposed Robust-NTC framework operates in an adaptive manner by leveraging the statistical parameters $(\boldsymbol{\mu},\boldsymbol{\sigma}^2)$. 
These parameters encapsulate the element-wise uncertainty and serve as control variables for both quantization and transmission decisions. 
Given $(\mu_i,\sigma_i^2)$, the transmitter selects the quantization configuration $(b_i,\boldsymbol{\epsilon}_i)$ and applies the corresponding channel-optimized quantizer such that the end-to-end distortion does not exceed the Gaussian-proxy target defined in~\eqref{eq:mmse_equiv}. 
Formally,
\begin{align}
D^{\star}(\sigma_i^2; b_i,\boldsymbol{\epsilon}_i)
\le \frac{\sigma_i^2}{\sigma_i^2+1},
\label{eq:dist-match-general}
\end{align}
which is equivalently expressed (for $\sigma_i^2>0$) as the normalized condition
\begin{align}
D^{\star}(1; b_i,\boldsymbol{\epsilon}_i)
\le \frac{1}{\sigma_i^2+1},
\label{eq:dist-match-unit}
\end{align}
defining the \emph{distortion-matching rule} for adaptive operation.
This rule assigns higher bit depths and more reliable channels to high-variance latents, while low-variance latents are encoded with fewer bits for efficiency, achieving distortion-consistent and rate-efficient communication across varying channel conditions.
By jointly adapting to source and channel statistics in real time, the proposed framework achieves scalable channel-aware JSCC without re-training.
The general applicability of the framework will be further demonstrated in the next section, where it is applied to a representative OFDM transmission model under practical conditions.

\vspace{2mm}
\noindent\textbf{Remark 1 (Relaxation for negligible-variance latents):}
The condition in~\eqref{eq:dist-match-unit} implies that the right-hand side is less than~1 for all $\sigma_i^2>0$, while $D^{\star}(1;0,\boldsymbol{\epsilon}_i)=1$ holds for any $\boldsymbol{\epsilon}_i$. 
As a result, the condition inherently enforces $b_i\!\ge\!1$ even for latents with negligible variance. 
To prevent redundant bit allocation for such elements, we relax this constraint by setting $b_i=0$ whenever $\sigma_i^2<\delta$, for a small threshold $\delta>0$. 
This relaxation is implicitly applied in all subsequent formulations.

\section{Application to Practical OFDM Systems: Robust-NTC over OFDM}\label{Sec:DigitalSC}
Building upon the Robust-NTC framework developed in the previous section, this section demonstrates its general applicability by integrating it into a practical OFDM system operating over frequency-selective fading channels.

\subsection{Problem Formulation}\label{Sec:OFDM}  
In this subsection, we specialize the general digital JSCC model of Sec.~\ref{Sec:System} to an OFDM transmission setting. 
In this case, the channel operator $\mathcal{W}(\cdot;\boldsymbol{\eta})$ is realized through OFDM modulation, resource mapping, subcarrier-dependent fading, and additive noise.  
As described in Sec.~\ref{Sec:System_model}, each latent element $y_i$ is quantized into a binary codeword $\mathbf{c}_i\in\{0,1\}^{b_i}$, and the overall transmit bitstream is expressed as $\mathbf{c}$.
This bitstream is then segmented, mapped onto modulation symbols, and assigned to resource elements (REs) in the OFDM time-frequency grid.
For OFDM symbol index $t\in\{1,\ldots,T_{\rm sym}\}$ and subcarrier index $k\in\{1,\ldots,N_{\rm sc}\}$, the modulated symbol is denoted as
\begin{align}
    s[t,k] \in \mathcal{X}_{2^m}, ~~ (t,k)\in \mathcal{I}_{\rm data},
\end{align}
where $\mathcal{I}_{\rm data}$ denotes the set of REs carrying data, and $\mathcal{X}_{2^m}$ is a $2^m$-QAM constellation drawn from the candidate set
\begin{align}
    \mathcal{X}_{\rm cand} = \{\mathcal{X}_4, \mathcal{X}_{16}, \mathcal{X}_{64}, \mathcal{X}_{256}\},
\end{align}
with each constellation normalized such that $\mathbb{E}[|s[t,k]|^2]=1$.  
Transmission power is adaptively allocated across subcarriers, with $p_{t,k}$ denoting the power allocated to subcarrier $k$ of the $t$-th OFDM symbol, under the constraint $\sum_{k=1}^{N_{\rm sc}} p_{t,k} \le P_{\rm tot}$.  
The modulated symbols are scaled accordingly,
\begin{align}
    {\tilde s}[t,k] = \sqrt{p_{t,k}}\,s[t,k],
\end{align}
which are converted into the time domain using an $N_{\rm sc}$-point IFFT and appended with a cyclic prefix (CP).
At the receiver, the CP is removed and an FFT is applied to reconstruct the frequency-domain received symbols. 
Let $h[t,k]\in\mathbb{C}$ denote the channel frequency response on subcarrier $k$ of the $t$-th OFDM symbol. 
Assuming that the channel remains constant over a coherence interval larger than the transmission block, i.e., $h[t,k]=h_k$ for all $t$, the received signal at subcarrier $k$ of symbol $t$ is expressed as
\begin{align}
    r[t,k] = \sqrt{p_{t,k}}\,h_k\,s[t,k] + v[t,k],
\end{align}
with $v[t,k]\sim\mathcal{CN}(0,\sigma^2)$.  
Perfect channel estimation is assumed at the receiver, and equalization is performed as
\begin{align}
    \hat{s}[t,k] &= (\sqrt{p_{t,k}}\,h_k)^{-1}r[t,k] \\
                 &= s[t,k] + (\sqrt{p_{t,k}}\,h_k)^{-1}v[t,k],
\end{align}
and the equalized symbols $\{\hat{s}[t,k]\}$ are demodulated into bits and reassembled into a noisy bitstream $\hat{\mathbf{c}}$, consistent with the general model $\hat{\mathbf{c}}=\mathcal{W}(\mathbf{c};\boldsymbol{\eta})$. 
 
\begin{figure}[t]
    \centering
    \includegraphics[width=9cm]{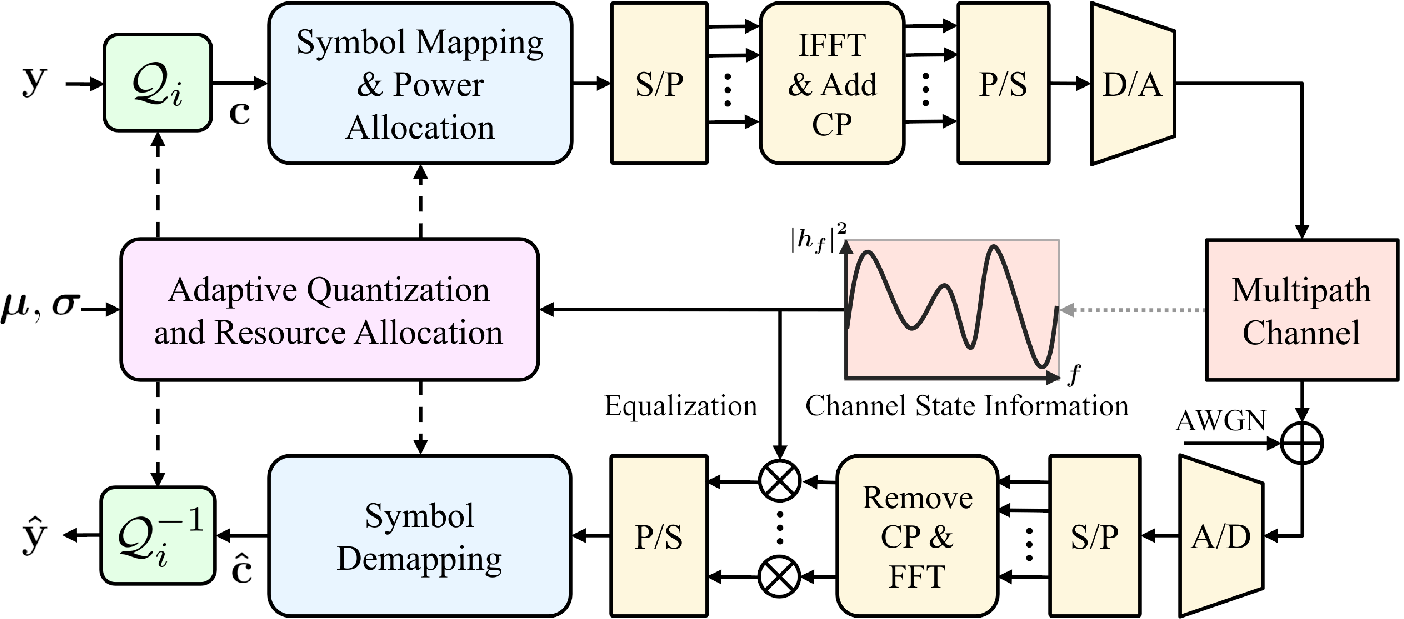}
    \caption{OFDM transmission model with the associated optimization framework.}
    \label{fig:OFDM_system}
\end{figure}

In this scenario, the objective of the Robust-NTC framework is to minimize the required number of OFDM symbols $T_{\rm sym}$ while satisfying the distortion-matching criterion established in Sec.~\ref{Sec:Adaptive_Inference}.
Let $\mathcal{Q}=\{{\mathcal{T}_i},{\mathcal R}_i\}_{i=1}^{N}$ denote the quantizer design, $\mathbf{b}=[b_1,\ldots,b_N]^{\sf T}$ denote the element-wise bit allocation vector, 
$\mathbf{P}\in\mathbb{R}^{T_{\rm sym}\times N_{\rm sc}}$ denote the power allocation matrix with $[\mathbf{P}]_{t,k} = p_{t,k}$, 
$\mathbf{M}\in\{0,2,4,6,8\}^{T_{\rm sym}\times N_{\rm sc}}$ denote the modulation order matrix such that $[\mathbf{M}]_{t,k} = m_{t,k}$, and $\boldsymbol \Pi$ denote the mapping from latent bits to OFDM REs.
The optimization problem is then formulated as
\begin{subequations}
\begin{align}
\min_{\mathcal{Q},\,\mathbf{b},\,\mathbf{P},\,\mathbf{M},\,\boldsymbol \Pi}
~~&  T_{\rm sym}  \\
\text{s.t.}\quad~~~~
& \mathbb{E}\!\left[(y_i-\hat{y}_i)^2\right]
   \le \frac{\sigma_i^2}{\sigma_i^2+1}, \quad \forall i,  \\
& \sum_{k=1}^{N_{\rm sc}} p_{t,k} \le P_{\rm tot}, \quad \forall t,  \\
& \sum_{i=1}^N b_i
   \leq \sum_{t=1}^{T_{\rm sym}}\sum_{k=1}^{N_{\rm sc}} m_{t,k},
\end{align}\label{eq:ofdm_formulation}
\end{subequations}
with the shared side information $(\boldsymbol{\mu},\boldsymbol{\sigma}^2)$ enabling coordinated resource mapping between transmitter and receiver. 
\textcolor{black}{The optimization problem \eqref{eq:ofdm_formulation} jointly captures source-side distortion requirements and channel-side transmission feasibility. Specifically, the distortion constraint ensures that each latent element $y_i$ is reconstructed with distortion no larger than the target level determined by $\sigma_i^2$, while the bit-budget constraint requires that the total number of allocated latent bits be fully supported by the modulation symbols placed on the OFDM time-frequency grid under the given power constraint.}
To explicitly relate this formulation to the BSC-based distortion model of Sec.~\ref{Sec:Analysis}, we approximate the subcarrier-level bit error probabilities.  
For Gray-labeled $2^{m_{t,k}}$-QAM, the BER on subcarrier $k$ of symbol $t$ is well approximated as~\cite{BER}
\begin{align}
    P_{b}(m_{t,k},\gamma_{t, k}) \approx&
    {\frac{4}{m_{t,k}}\!\left(1-\frac{1}{\sqrt{2^{m_{t,k}}}}\right)
    {\mathrm Q}\left(\sqrt{\frac{3\gamma_{t, k}}{2^{m_{t,k}}-1}}\right)} \nonumber \\
    & + \frac{4}{m_{t,k}}\!\left(1-\frac{2}{\sqrt{2^{m_{t,k}}}}\right)
    {\mathrm Q}\left(3\sqrt{\frac{3\gamma_{t, k}}{2^{m_{t,k}}-1}}\right) \nonumber \\
    \triangleq& \hat{P}_{b}(m_{t,k},\gamma_{t, k}),
    \label{eq:qam_ber_ofdm}
\end{align}
where $\gamma_{t, k}=p_{t,k}|h_k|^2/\sigma^2$, and ${\mathrm Q}(\cdot)$ denotes the Gaussian $Q$-function.  
These estimated error probabilities define the effective flip probabilities $\hat{\boldsymbol{\epsilon}}_i$ experienced by the $b_i$ bits of latent codeword $\mathbf{c}_i$, determined by the mapping $\boldsymbol \Pi$.  
Conditioned on the knowledge of $\{\hat{\boldsymbol{\epsilon}}_i\}_{i=1}^{N}$ and under the assumption that each quantizer is optimized for the corresponding $(b_i,\hat{\boldsymbol{\epsilon}}_i)$ pair, the optimization problem can be reformulated as
\begin{subequations}\label{eq:ofdm_formulation_BSC}
\begin{align}
\min_{\mathbf{b},\,\mathbf{P},\,\mathbf{M},\,\boldsymbol \Pi}
~~&  T_{\rm sym}  \\
\text{s.t.}\quad~~
& D^{\star}(1; b_i,\hat{\boldsymbol{\epsilon}}_i)
   \le \frac{1}{\sigma_i^2+1}, \quad \forall i,  \\
& \sum_{k=1}^{N_{\rm sc}} p_{t,k} \le P_{\rm tot}, \quad \forall t,  \\
& \sum_{i=1}^N b_i
   \leq \sum_{t=1}^{T_{\rm sym}}\sum_{k=1}^{N_{\rm sc}} m_{t,k}.
\end{align}
\end{subequations}
This reformulation provides an explicit bridge between the physical-layer OFDM parameters and our distortion-matching criterion.
The overall transceiver chain, together with the associated optimization framework, is summarized in Fig.~\ref{fig:OFDM_system}.

\subsection{Optimization for Efficient Resource Allocation}\label{Sec:Transmit_OFDM}
The reformulated optimization problem in \eqref{eq:ofdm_formulation_BSC} jointly involves bit allocation, bit mapping, power allocation, and modulation selection. 
Due to the discrete nature of the design variables, the problem is non-convex and combinatorial, rendering a direct solution intractable.
Moreover, obtaining the element-wise optimal quantizer for each $(b_i,\boldsymbol{\epsilon}_i)$ pair requires prohibitively high computational complexity.

To circumvent this challenge, we first construct an offline quantizer library for a finite set of BER targets $\mathcal{E}=\{\bar\epsilon^{(1)},\ldots,\bar\epsilon^{(Q)}\}$.
Specifically, we define
\begin{align}
\! \! \mathcal{Q}_{\rm Lib}
= \Big\{\mathcal{Q}^{(b,q)} :
b\in\{1,\ldots,b_{\max}\},~
q\in\{1,\ldots,Q\}\Big\},
\end{align}
where each entry \(\mathcal{Q}^{(b,q)}\) stores the quantizer optimized for \((b,\bar\epsilon^{(q)})\) with its normalized distortion \(D^{\star}(1;b,\bar{\boldsymbol \epsilon}^{(q)})\).
Subsequently, we impose an additional constraint that enforces a common flip probability across all subcarriers by selecting a single BER target \(\bar\epsilon\in\mathcal{E}\), thereby removing the need to optimize $\boldsymbol{\Pi}$. 
With these constraints, the optimization problem can be rewritten as
\begin{subequations}\label{eq:ofdm_formulation_eq_BSC}
\begin{align}
\min_{\mathbf{b},\,\mathbf{P},\,\mathbf{M},\,  \bar\epsilon}
~~&  T_{\rm sym}  \\
\text{s.t.}\quad~~ 
& D^{\star}(1; b_i,\boldsymbol{\bar \epsilon}_i)
   \le \frac{1}{\sigma_i^2+1}, \quad \forall i,  \\
& \sum_{k=1}^{N_{\rm sc}} p_{t,k} \le P_{\rm tot}, \quad \forall t,  \\
& \sum_{i=1}^N b_i
   \leq \sum_{t=1}^{T_{\rm sym}}\sum_{k=1}^{N_{\rm sc}} m_{t,k},  \\
&\hat{P}_{b}(m_{t,k},\gamma_{t, k}) = \bar\epsilon, \quad \forall t,k, \label{eq:bsc_matching} \\ 
&\bar\epsilon \in \mathcal{E},
\end{align}
\end{subequations}
where $\boldsymbol{\bar \epsilon}_i = [\bar \epsilon, \ldots, \bar \epsilon]^{\sf T} \in [0,0.5]^{b_i}$.
The constraint in \eqref{eq:bsc_matching} is equivalently expressed by determining the transmit power that achieves the target BER $\bar\epsilon$ for each subcarrier, given its modulation order $m_{t,k}$ and channel gain $h_k$.
Specifically, the required power is obtained as
\begin{align}\label{eq:power_replaced}
p_{t,k}(m_{t,k},\bar\epsilon)
= \frac{\gamma_{\rm th}(m_{t,k},\bar\epsilon)\sigma^2}{|h_k|^2},
\end{align}
where $\gamma_{\rm th}(m,\bar\epsilon)$ denotes the SNR threshold that meets $\hat{P}_{b}(m,\gamma_{\rm th})=\bar\epsilon$, precomputed through a bisection search exploiting the monotonicity of the BER function.

To solve \eqref{eq:ofdm_formulation_eq_BSC}, we optimize the problem separately for each candidate BER target $\bar\epsilon \in \mathcal{E}$ and then select the one that yields the smallest $T_{\rm sym}$.
For a fixed $\bar\epsilon$, the problem decomposes into two subproblems.
The first subproblem focuses on the source side: under a fixed $\bar\epsilon$, the distortion of each latent dimension depends solely on the allocated bit depth $b_i$. Consequently, the distortion-matching condition reduces to the following bit-allocation problem:
\begin{subequations}\label{eq:R_lat_fixed}
\begin{align}
 ({\mathbf P1})~~\min_{\mathbf{b}}
~~& B_{\rm lat}(\bar\epsilon) \triangleq \sum_{i=1}^N b_i  \\
\text{s.t.}~~
& D^{\star}(1; b_i,\boldsymbol{\bar \epsilon}_i)
   \le \frac{1}{\sigma_i^2+1}, \quad \forall i.
\end{align}
\end{subequations}
The above problem $({\mathbf{P1}})$ is readily solved by:
\begin{align}
b_i^{\star}
= \min\Big\{b_i \in {1,\ldots,b_{\max}}:
D^{\star}(1; b_i, \bar{\boldsymbol \epsilon}) \le \frac{1}{\sigma_i^2+1}\Big\}.
\label{eq:bit_alloc_solution}
\end{align}
Second, under a fixed $\bar\epsilon$ and $B_{\rm lat}^\star(\bar\epsilon)$, the objective of minimizing $T_{\rm sym}$ is achieved by maximizing the number of bits transmittable per OFDM symbol while satisfying the BER target $\bar\epsilon$.
This leads to the following power and modulation optimization problem defined for a single OFDM symbol:
\begin{subequations}\label{eq:R_sym_fixed}
\begin{align}
({\mathbf P2})~~\max_{\mathbf{p},\,\mathbf{m}}
& \quad R_{\rm sym}(\bar\epsilon) \triangleq \sum_{k=1}^{N_{\rm sc}} m_{k}  \\
\text{s.t.}\quad
& \sum_{k=1}^{N_{\rm sc}} \frac{\gamma_{\rm th}(m_{k},\bar\epsilon)\sigma^2}{|h_k|^2} \le P_{\rm tot},
\end{align}
\end{subequations}
where the power allocation is replaced based on~\eqref{eq:power_replaced}.
The optimized allocation $(\mathbf{p}^\star(\bar \epsilon),\mathbf{m}^\star(\bar \epsilon))$ can then be applied identically across all $T_{\rm sym}$ symbols to construct $(\mathbf{P}^\star(\bar \epsilon),\mathbf{M}^\star(\bar \epsilon))$.
The problem $(\mathbf{P2})$ can be efficiently solved using a greedy algorithm~\cite{DiscreteOPT}, which successively increases the modulation order of the subcarrier that requires the minimum additional power per rate increment, as summarized in {\bf Algorithm~1}.

\begin{algorithm}[t]
{\small \caption{Power-Modulation Allocation for $(\mathbf{P2})$}
\label{alg:Alg_for_P2}
\begin{algorithmic}[1]
\REQUIRE Total power budget \(P_{\rm tot}\), channel gains \(\{|h_{k}|^2\}_{k=1}^{N_{\rm sc}}\), and SNR thresholds $\gamma_{\rm th}(m,\bar\epsilon),~\forall m$.
\ENSURE Modulation-power pairs $\{m_{k}, p_{k}\}_{k=1}^{N_{\rm sc}}$.
\STATE Set $\bar{P}=0$ and \(m_k = 0,~p_k = 0,~\forall k\).
\WHILE{true}
    \STATE Compute $\Delta p_{k}(m_{k})= \frac{\big(\gamma_{\rm th}(m_{k}+2,\bar\epsilon)-\gamma_{\rm th}(m_{k},\bar\epsilon)\big)\sigma^2}{|h_k|^2}$, $~\forall k$.
    \STATE Select $k^\star =\arg\!\min_{k} \Delta p_{k}(m_{k})$.
    \IF{$\bar{P} + \Delta p_{k^\star}(m_{k^\star}) \leq P_{\rm tot}$}
        \STATE Update $\bar{P} \gets \bar{P} + \Delta p_{k^\star}(m_{k^\star})$, $m_{k^\star} \gets m_{k^\star} + 2$.
        \STATE Update $p_{k^\star} = \frac{\gamma_{\rm th}(m_{k^{\star}},\bar\epsilon)\sigma^2}{|h_{k^{\star}}|^2}$.
    \ELSE
        \STATE Break the loop.
    \ENDIF
\ENDWHILE
\end{algorithmic}}
\end{algorithm}

Each candidate $\bar\epsilon\in\mathcal{E}$ yields a pair $(B_{\rm lat}^\star(\bar\epsilon), R_{\rm sym}^\star(\bar\epsilon))$. 
\textcolor{black}{Specifically, $(\mathbf{P1})$ determines the required latent bit budget to satisfy the distortion constraints, while $(\mathbf{P2})$ provides the achievable per-symbol throughput under the target BER and power constraint. The optimal operating point is then obtained by minimizing the ratio between these two quantities, since this ratio determines the required number of OFDM symbols:}
\begin{align}
\bar\epsilon^{\star}
= \underset{\bar\epsilon\in\mathcal{E}}{\arg\!\min}~
\frac{B_{\rm lat}^{\star}(\bar\epsilon)}{R_{\rm sym}^{\star}(\bar\epsilon)},
\label{eq:ratio_selection}
\end{align}
from which the minimum number of OFDM symbols is obtained as 
$T_{\rm sym}^{\star} = \lceil B_{\rm lat}^{\star}(\bar\epsilon^{\star}) / R_{\rm sym}^{\star}(\bar\epsilon^{\star}) \rceil$.
In practice, the optimized channel throughput is slightly higher than the required latent transmission load, i.e., $B_{\rm lat}^{\star}(\bar\epsilon^{\star})< T_{\rm sym}^{\star}R_{\rm sym}^{\star}(\bar\epsilon^{\star})$, leaving a small number of unused bits within the OFDM resource grid.

\begin{algorithm}[t]
{\small  \caption{Bit Refinement for $(\mathbf{P3})$}
\label{alg:Alg_for_P3}
\begin{algorithmic}[1]
\REQUIRE Optimized bit allocation $\mathbf{b}^\star$, residual bit budget $b_{\rm res} = T_{\rm sym}^\star R_{\rm sym}^\star(\bar\epsilon^\star) - B_{\rm lat}^{\star}(\bar\epsilon^{\star})$, distortion table $\{{D^{\star}(1;b, \bar{\boldsymbol \epsilon}^\star)}\}_{b=1}^{b_{\rm max}}$, and latent variances $\{\sigma_i^2\}_{i=1}^{N}$.
\ENSURE Refined bit allocation $\tilde{\mathbf{b}}$.
\STATE Set $\tilde{\mathbf{b}} = \mathbf{b}^\star$.
\WHILE{$b_{\rm res} > 0$}
    \STATE Compute $\Delta_i = \sigma_i^2\big(D^{\star}(1;\tilde b_i,\bar{\boldsymbol \epsilon}^\star) - D^{\star}(1;\tilde b_i{+}1,\bar{\boldsymbol \epsilon}^\star)\big)$, $\forall i$.
    \STATE Select $i^\star = \arg\!\max_i \Delta_i$.
    \STATE Update $\tilde b_{i^\star} \gets \tilde b_{i^\star} + 1$, $b_{\rm res} \gets b_{\rm res} - 1$ 
\ENDWHILE
\end{algorithmic}}
\end{algorithm}

To fully exploit this residual throughput, we perform an additional greedy bit refinement stage that incrementally allocates the remaining bits to minimize the total distortion across all latent dimensions, given that the distortion-matching condition is already satisfied.
This refinement is formulated as the following optimization problem:
\begin{subequations}\label{eq:B_greedy}
\begin{align}
 ({\mathbf P3})~~\min_{\mathbf{b}}
~~&\sum_{i=1}^{N}\sigma_i^2D^{\star}(1; b_i,\boldsymbol{\bar \epsilon}_i^\star) \\
\text{s.t.}~~
& D^{\star}(1; b_i,\boldsymbol{\bar \epsilon}_i^\star)
   \le \frac{1}{\sigma_i^2+1}, \quad \forall i. \\
&  \sum_{i=1}^N b_i = T_{\rm sym}^{\star}R_{\rm sym}^{\star}(\bar\epsilon^{\star}).
\end{align}
\end{subequations}
The problem $({\mathbf P3})$ can likewise be solved using a greedy algorithm, which iteratively allocates the remaining bits to the latent element yielding the maximum marginal distortion reduction, as summarized in {\bf Algorithm~2}.
Finally, the optimized quantizer, bit allocation, and transmission mapping corresponding to $\bar\epsilon^{\star}$ are applied to construct the final OFDM transmission configuration. 
The overall transmission optimization procedure is summarized in {\bf Algorithm 3}.

\vspace{1mm}

\noindent\textbf{Remark 2 (Bounding latent variance for consistent training and inference):}
Even when the maximum bit depth $b_{\max}$ is allocated, certain latent elements with excessively large variance $\sigma_i^2$ may not be guaranteed to meet the distortion-matching criterion in~\eqref{eq:dist-match-unit}, potentially leading to inconsistency between training and inference. 
In addition, an excessively large $b_{\max}$ increases optimization complexity and quantizer storage overhead.
To address these issues, a scaling constraint is incorporated from the training stage by applying an activation that bounds the predicted variance within a feasible range:
\begin{align}
\sigma_i
= \sigma_{\max}\tanh\!\left(
\frac{\mathrm{softplus}(\tilde\sigma_i)}{\beta}
\right),
\end{align}
where $\beta$ is a learnable temperature parameter, and
\begin{align}
\sigma_{\max}
= \sqrt{\frac{1}{D^{\star}(1; b_{\max},\boldsymbol{\bar\epsilon}_i^{(\max)})}-1},
\end{align}
with $\bar\epsilon_i^{(\max)}$ denoting the largest element in $\mathcal{E}$.  
To further suppress distributional outliers, latent elements exceeding three standard deviations (i.e., $|y_i-\mu_i|>3\sigma_i$) are clipped to the boundary, and the corresponding gradients are learned through a straight-through estimator (STE)~\cite{STE}.

\vspace{1mm}

\noindent\textbf{Remark 3 (Optimality of Algorithms 1 and 2):}
The optimality of the proposed greedy algorithms follows from the convexity of their respective cost functions~\cite{DiscreteOPT}. 
For $(\mathbf{P2})$, the required SNR $\gamma_{\mathrm{th}}(m,\bar\epsilon)$ exhibits an increasing and convex trend with respect to the modulation order $m$, ensuring that the greedy allocation achieves the global optimum.
Similarly, for $(\mathbf{P3})$, the distortion $D^{\star}(1;b,\bar{\boldsymbol\epsilon}^\star)$ shows a decreasing and convex trend with respect to the bit depth $b$, as empirically observed in Fig.~\ref{fig:quantizer_opt_merge}, thereby confirming the optimality of the refinement process.

\vspace{1mm}

\begin{algorithm}[t]
{\small \caption{Overall Optimization Procedure for Robust-NTC over OFDM}
\label{alg:Overall_RobustNTC}
\begin{algorithmic}[1]
\REQUIRE Total power budget $P_{\rm tot}$, channel gains $\{|h_k|^2\}_{k=1}^{N_{\rm sc}}$, quantizer library $\mathcal{Q}_{\rm Lib}$, latent variances $\{\sigma_i^2\}_{i=1}^{N}$, and SNR thresholds $\gamma_{\rm th}(m,\bar\epsilon),~ \forall (m,\bar \epsilon)$.
\ENSURE Final operating point $\bar\epsilon^{\star}$, optimized parameters 
$(\mathcal{Q}^{\star}, \mathbf{b}^{\star}, \mathbf{P}^{\star}, \mathbf{M}^{\star})$.
\FOR{each $\bar\epsilon^{(q)} \in \mathcal{E}$}
    \STATE Solve $({\mathbf P1})$ to obtain $B_{\rm lat}^{\star}(\bar\epsilon^{(q)})$.
    \STATE Solve $({\mathbf P2})$ using {\bf Algorithm 1} 
           to obtain $R_{\rm sym}^{\star}(\bar\epsilon^{(q)})$.
\ENDFOR
\STATE Select $\bar\epsilon^{\star} = \arg\!\min_{\bar\epsilon^{(q)} \in \mathcal{E}} B_{\rm lat}^{\star}(\bar\epsilon^{(q)}) / R_{\rm sym}^{\star}(\bar\epsilon^{(q)})$.
\STATE Compute $T_{\rm sym}^{\star} = \lceil B_{\rm lat}^{\star}(\bar\epsilon^{(q)}) / R_{\rm sym}^{\star}(\bar\epsilon^{(q)}) \rceil$.
\IF{$T_{\rm sym}^{\star}R_{\rm sym}^{\star}(\bar\epsilon^{\star}) > B_{\rm lat}^{\star}(\bar\epsilon^{\star})$}
    \STATE Refine $\mathbf{b}^{\star}$ using {\bf Algorithm 2}.
\ENDIF
\STATE Construct $(\mathcal{Q}^{\star}, \mathbf{b}^{\star}, 
                  \mathbf{P}^{\star}(\bar\epsilon^{\star}),
                  \mathbf{M}^{\star}(\bar\epsilon^{\star}))$ 
                  and map bits to the OFDM resource grid.
\end{algorithmic}}
\end{algorithm}

\noindent\textbf{\textcolor{black}{Remark 4 (Computational complexity and latency mitigation):}}
    \textcolor{black}{The computational complexity of \textbf{Algorithm 3} can be expressed as $\mathcal{O}(Q  N  b_{\rm max} + Q  N_{sc}  b_{\rm power} + N  b_{\rm res})$, where $Q$, $b_{\rm power}$, and $b_{\rm res}$ denote the number of candidate BER targets, the number of greedy power-allocation updates, and the number of residual-bit allocations, respectively. In practice, the processing latency is mainly determined by the degree of sequential processing. While the evaluation of the $Q$ candidates and the solution of subproblem {\bf (P1)} via precomputed tables are highly parallelizable, the strict greedy updates in \textbf{Algorithms 1 and 2} constitute the main latency bottleneck. A practical way to alleviate this burden is to relax the strict greedy procedure by selecting multiple efficient candidate increments at each iteration, which reduces the number of sequential update rounds at the cost of only a minor deviation from the exact greedy path. In addition, the complexity can be further reduced through subcarrier grouping and by reusing the same power-modulation allocation over the channel coherence interval, rather than recomputing it for every transmitted image.}
\vspace{1mm}

\noindent\textbf{\textcolor{black}{Remark 5 (Extension to other channel and system models):}}
\textcolor{black}{While this paper primarily focuses on the OFDM setting, the proposed Robust-NTC framework can, in principle, be extended to other channel and system models. This can be achieved by redefining the distortion measure to reflect the specific impairments of the target environment and then optimizing the associated transmission parameters to satisfy the corresponding distortion-matching rule. For example, in multiple-input multiple-output (MIMO) systems~\cite{MIMO_JSCC}, the framework can be extended by treating spatial streams as parallel transmission dimensions. Likewise, in interference-limited scenarios such as multi-user broadcast or interference channels~\cite{Multi_user_MIMO_JSCC}, the current SNR-based analysis can be generalized to a signal-to-interference-plus-noise ratio (SINR)-based one. In such extensions, the resource-allocation optimization should be carefully tailored to strike an appropriate balance between computational complexity and optimality.}

\section{Simulation Results}\label{sec:Simulation}

\begin{table*}[t]
\caption{Network Configurations of the Robust-NTC Framework.}
\label{tab:network_config}
\centering
\footnotesize
\renewcommand{\arraystretch}{1.08}
\begin{tabular}{@{}lccc@{}}
\toprule
\textbf{Component Description} & \textbf{CIFAR-10} & \textbf{STL-10} & \textcolor{black}{\textbf{Kodak (ImageNet)}} \\
\midrule

\multicolumn{4}{@{}l}{\textbf{Swin-Transformer Analysis \& Synthesis Networks ($g_a$, $g_s$)}} \\
\quad Stages / Window size & (2, 8) & (3, 6) & \textcolor{black}{(4, 8)} \\
\quad Embedding dimensions ($C$) per stage & [64, 64] & [192, 192, 192] & \textcolor{black}{[256, 256, 256, 256]} \\
\quad Number of attention heads per stage & [8, 8] & [8, 8, 8] & \textcolor{black}{[8, 8, 8, 8]} \\
\quad Number of Swin blocks per stage & [2, 2] & [2, 2, 2] & \textcolor{black}{\begin{tabular}[c]{@{}c@{}}[1, 1, 2, 4] (reversed for $g_s$)\end{tabular}} \\
\midrule

\multicolumn{4}{@{}l}{\textbf{CNN-based Hyperprior Networks ($h_a$, $h_s$)}} \\
\quad Hyper-encoder ($h_a$) feature channels & [64, 48, 48, 48] & [192, 96, 96, 96] & \textcolor{black}{[256, 192, 192, 192]} \\
\quad Hyper-decoder ($h_s$) feature channels & [48, 64, 96, 128] & [96, 192, 288, 384] & \textcolor{black}{[192, 256, 384, 512]} \\
\quad Convolution kernel size ($K$) \& stride ($S$) & \multicolumn{3}{c}{$(3,1), (5,2), (5,2)$} \\
\quad Activation functions & \multicolumn{3}{c}{LeakyReLU for $h_a$, ReLU for $h_s$} \\
\midrule

\multicolumn{4}{@{}l}{\textbf{Latent Refinement Network (MLP)}} \\
\quad Hidden layer dimensions & [192, 384, 64] & [576, 1152, 192] & \textcolor{black}{[768, 1536, 256]} \\
\quad Activation function & \multicolumn{3}{c}{Gaussian Error Linear Unit (GELU)} \\
\bottomrule
\end{tabular}
\end{table*}

This section evaluates the reconstruction fidelity and transmission efficiency of the proposed Robust-NTC framework for wireless image transmission in OFDM systems. 
The transmission channel follows the 3GPP TDL-C model \cite{3GPP_TR_38_901_v18} with a delay spread of $300$~ns at a carrier frequency of $3.5$~GHz. 
The OFDM system employs $N_{\mathrm{sc}}=512$ subcarriers with a subcarrier spacing of $\Delta f=30$~kHz. 
The total transmission power is defined as $P_{\mathrm{tot}}\!=\!N_{\mathrm{sc}}P$, 
and the SNR is given by $P/\sigma^2$.
The quantizer library is constructed for ten channel conditions, with target BERs uniformly sampled in the logarithmic (dB) domain over $\epsilon \in [0.001,0.05]$.
The maximum quantization bit depth and negligible-variance threshold are set to $b_{\max}=8$ and $\delta=0.4$, respectively.

\textcolor{black}{The simulations are conducted using the CIFAR-10, STL-10, and Kodak datasets.
The CIFAR-10 dataset contains $60{,}000$ color images of size $3\times 32\times32$ (50{,}000 for training and 10{,}000 for testing), 
while STL-10 comprises $113{,}000$ images of size $3\times96\times96$ (105{,}000 training and 8{,}000 test samples). 
For higher-resolution image reconstruction, the proposed framework is additionally evaluated on the Kodak dataset. Specifically, training is performed using 50{,}000 images sampled from the ImageNet validation set \cite{ImageNet}, where random $3\times 256 \times 256$ patches are extracted during training. Evaluation is then conducted on the Kodak dataset, which contains 24 images of size $3\times768\times512$.}
Reconstruction performance is evaluated in terms of the peak signal-to-noise ratio (PSNR), defined as
\begin{align}
\mathrm{PSNR}=10\log_{10}\!\left(\frac{255^2}{\mathrm{MSE}}\right),
\end{align}
with $\mathrm{MSE}=\tfrac{1}{CHW}\|\mathbf{x}-\hat{\mathbf{x}}\|_2^2$. 
All models, including the baselines, are trained using the Adam optimizer with a learning rate of either $10^{-4}$ or $10^{-3}$ for a sufficient number of iterations to achieve stable performance.
The model follows the architecture in~\cite{NTSCC}, comprising a Swin-Transformer-based analysis and synthesis network~\cite{Swin_trans}, 
along with convolutional neural network (CNN) based hyper-analysis and hyper-synthesis transforms \cite{NTC}. 
For CIFAR-10, a compact two-stage Swin configuration is used for efficiency, while STL-10 adopts a deeper three-stage variant with larger embeddings.
Detailed network configurations are summarized in Table~\ref{tab:network_config}.

\begin{table}[t]
\caption{Average number of side-information bits per image.}
\label{tab:side_info_bits}
\centering
\footnotesize
\renewcommand{\arraystretch}{1.08}
\setlength{\tabcolsep}{4pt}
\begin{tabular}{lccc}
\toprule
\multirow{2}{*}{\textbf{Scheme}} &
\multicolumn{3}{c}{\textbf{Side bits [bits/image]}} \\
\cmidrule(lr){2-4}
& \textbf{CIFAR-10} & \textbf{STL-10} & \textbf{\textcolor{black}{Kodak}} \\
\midrule
Robust-NTC ($\lambda_y{=}192, \lambda_z{=}192$) & 46.34 & 401.14 & -- \\
Robust-NTC ($\lambda_y{=}128, \lambda_z{=}128$) & 78.01 & 462.97 & -- \\
\textcolor{black}{Robust-NTC ($\lambda_y{=}512, \lambda_z{=}512/0.75$)} & \textcolor{black}{--} & \textcolor{black}{--} & \textcolor{black}{6150.73} \\
\textcolor{black}{Robust-NTC ($\lambda_y{=}768, \lambda_z{=}768/0.5$)} & \textcolor{black}{--} & \textcolor{black}{--} & \textcolor{black}{3652.01} \\
MDJCM-$M$QAM (Equal / Opt) & 256.00 & 556.00 & \textcolor{black}{6144.00} \\
\bottomrule
\end{tabular}
\end{table}

To evaluate the proposed Robust-NTC framework, several digital JSCC baselines are considered.
\begin{itemize}
\item \textbf{E2E-Fixed-$M$QAM:}
This method jointly trains an end-to-end JSCC model with an OFDM transceiver using a STE~\cite{STE} for differentiable QAM modulation and demodulation with fixed OFDM symbol length. 
The transmitter employs an MLP-based refinement module that takes both the channel response and latent vector as inputs before QAM mapping, while power allocation follows the theoretical capacity-achieving waterfilling strategy~\cite{Waterfilling}. 
Each model is trained separately for a specific SNR and channel distribution under a fixed transceiver configuration, resulting in limited scalability and poor generalization to unseen conditions.

\item \textbf{Fixed-$b$-Quant-$\epsilon$~\cite{OFDM_ICT}:}
This method employs a fixed $b$-bit uniform scalar quantizer trained under a single bit-flip probability $\epsilon$. 
Given the trained $\epsilon$, Algorithm~1 in Sec.~\ref{Sec:Transmit_OFDM} is applied to allocate power and modulation levels satisfying the target BER constraint.
For a given parameter pair $(b,\epsilon)$, a single trained model is employed for all evaluation scenarios.

\item \textcolor{black}{\textbf{MDJCM-$M$QAM~\cite{MDJCM}:}
This method integrates a deepJSCC transmission module into the NTC framework~\cite{NTSCC} with digital modulation. 
The model employs an adaptive transmission module that takes both the SNR and modulation order as inputs. For OFDM-based evaluation, two variants are implemented. The first, \textbf{MDJCM-Equal-$M$QAM}, allocates power so that all active subcarriers yield the same effective SNR, and the resulting common effective SNR is provided to the adaptive transmission module. The second, \textbf{MDJCM-Opt-$M$QAM}, excludes severely attenuated subcarriers to reduce the influence of weak channels by determining the optimal number of active subcarriers that maximizes capacity under the equal effective SNR constraint. Specifically, after sorting the subcarriers in descending order of channel gain, the optimal number of active subcarriers $K^\star \le N_{\mathrm{sc}}$ is given by:
\begin{equation}
    K^\star = \argmax_{K \in \{1, \dots, N_{\mathrm{sc}}\}} K \log_2\big(1 + \gamma_{\mathrm{eff}}(K)\big),
\end{equation}
where $\gamma_{\mathrm{eff}}(K)$ denotes the common effective SNR when the total transmission power is distributed over the top $K$ subcarriers. Unlike the E2E-$M$QAM baseline, a single MDJCM model trained by sampling over diverse channel conditions is used for all evaluation scenarios.}

\begin{figure}[t]
    \centering
    \includegraphics[width=0.8\columnwidth]{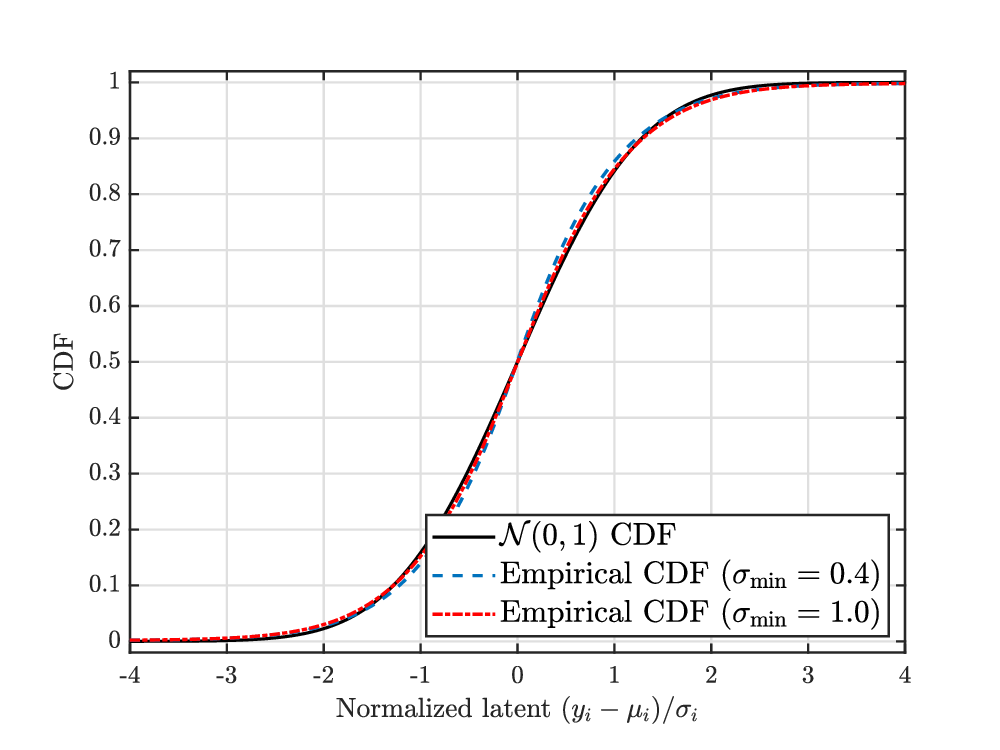}
    \caption{Empirical CDFs of normalized latents $(y_i-\mu_i)/\sigma_i$ obtained from the CIFAR-10 test dataset using the model learned under the proposed Robust-NTC framework, compared with the standard Gaussian reference $\mathcal{N}(0,1)$.}
    \label{fig:Latent_CDF}
\end{figure}

\begin{figure*}[t]
    \centering 
    \begin{minipage}{2\columnwidth}
        \centering 
        {\setlength{\fboxrule}{0pt}
        \subfigure[$\mathrm{SNR}=5$ dB]
         {\epsfig{file=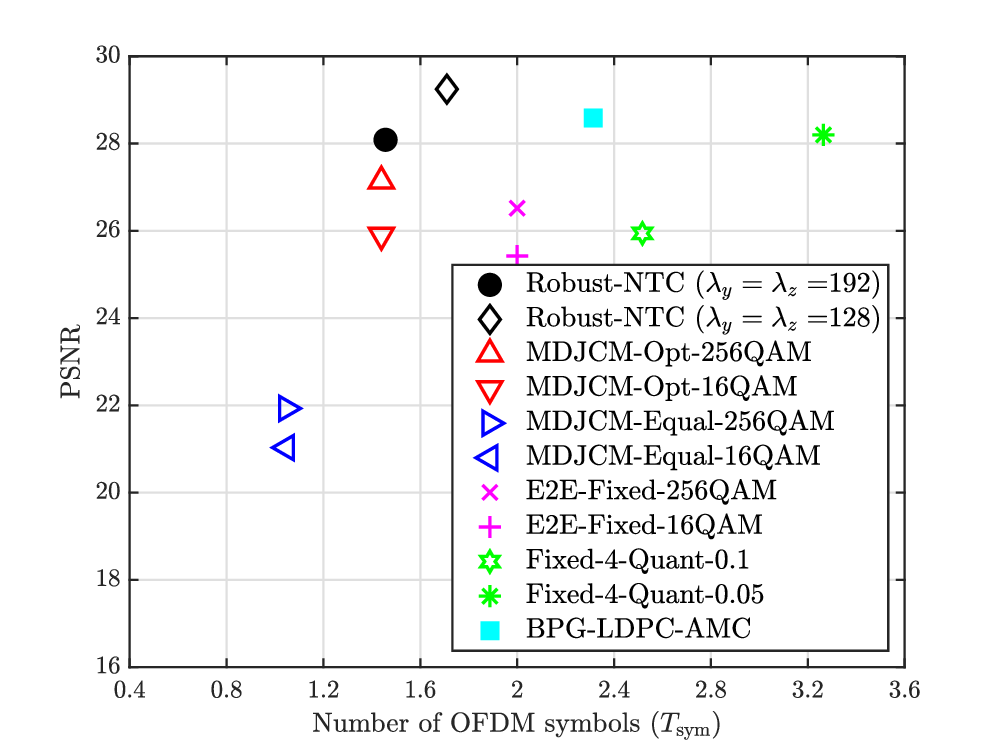, width=6.5cm}}\hspace*{-6mm}
        \subfigure[$\mathrm{SNR}=10$ dB]
    {\epsfig{file=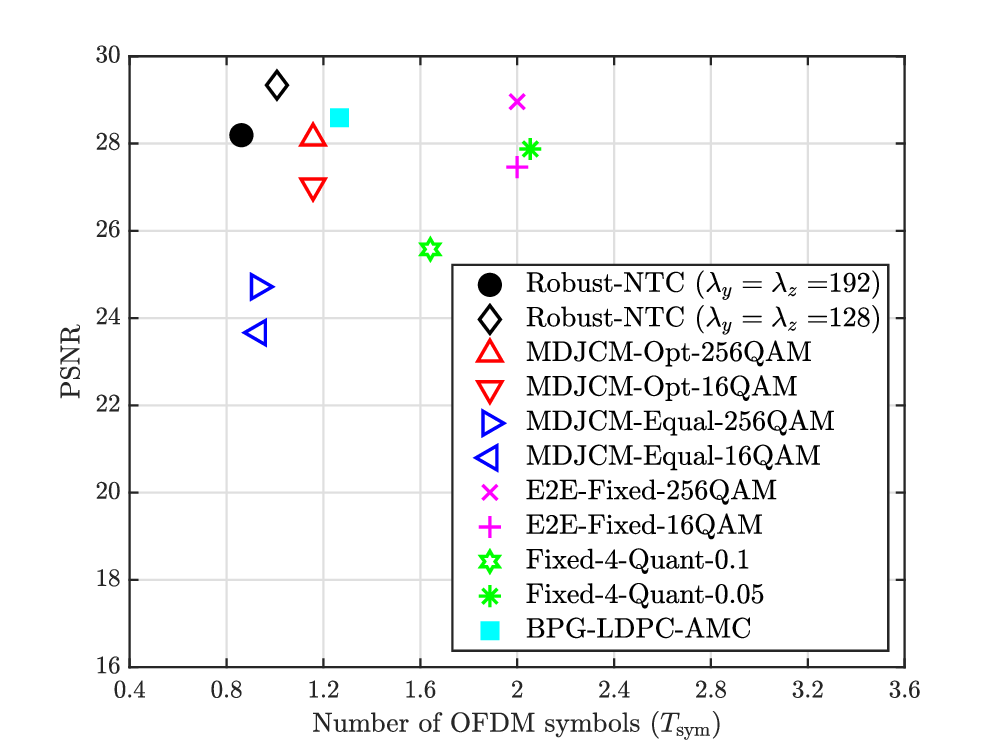, width=6.5cm}}\hspace*{-6mm}
        \subfigure[$\mathrm{SNR}=15$ dB]
    {\epsfig{file=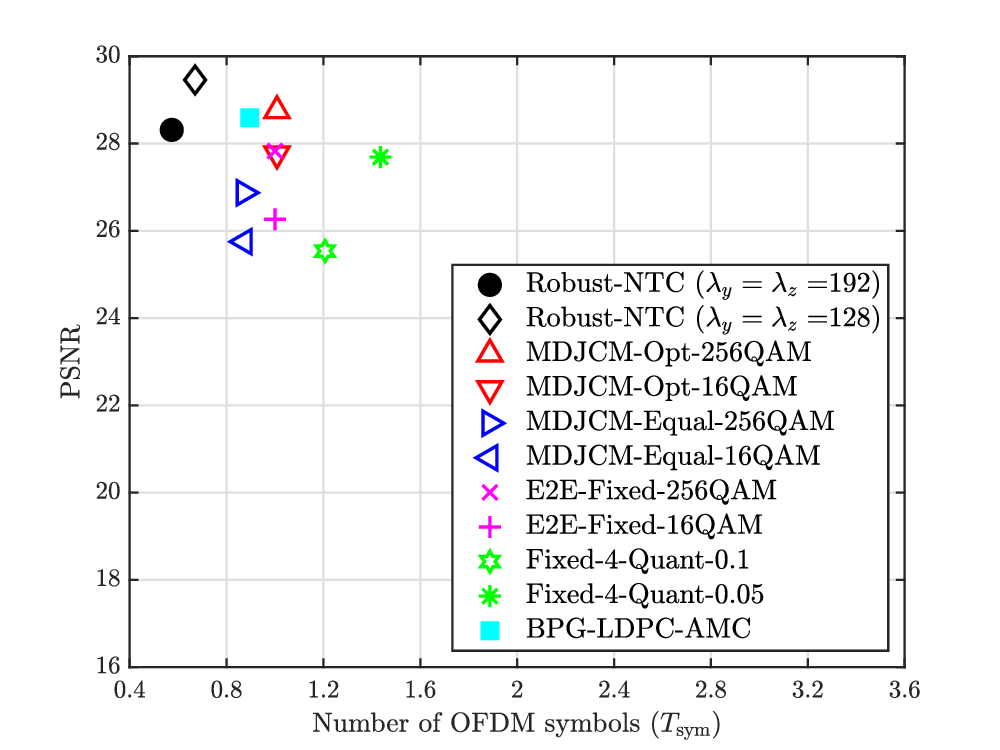, width=6.5cm}}\hspace*{-6mm}
        \caption{\textcolor{black}{Comparison of the proposed Robust-NTC and baseline schemes in terms of PSNR versus the number of OFDM symbols under different SNR levels, evaluated on the CIFAR-10 dataset.}}
        \label{fig:diff_SNR_PSNR}}
    \end{minipage}
\end{figure*}

\begin{figure*}[t]
    \centering 
    \begin{minipage}{2\columnwidth}
        \centering 
        {\setlength{\fboxrule}{0pt}
        \subfigure[$\mathrm{SNR}=5$ dB]
         {\epsfig{file=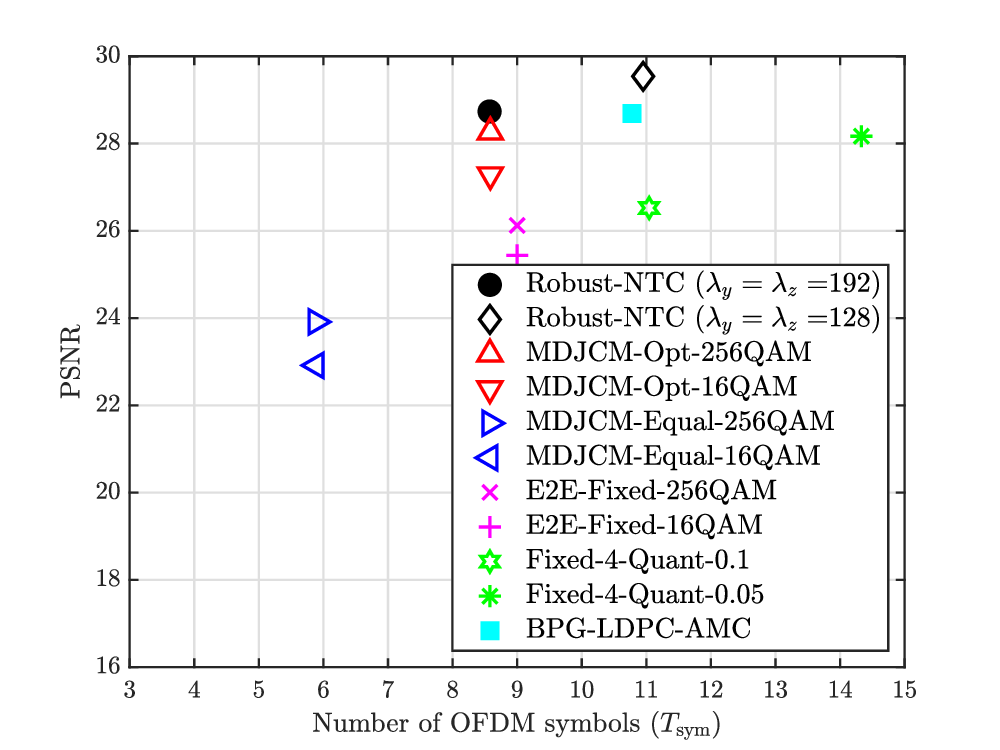, width=6.5cm}}\hspace*{-6mm}
        \subfigure[$\mathrm{SNR}=10$ dB]
    {\epsfig{file=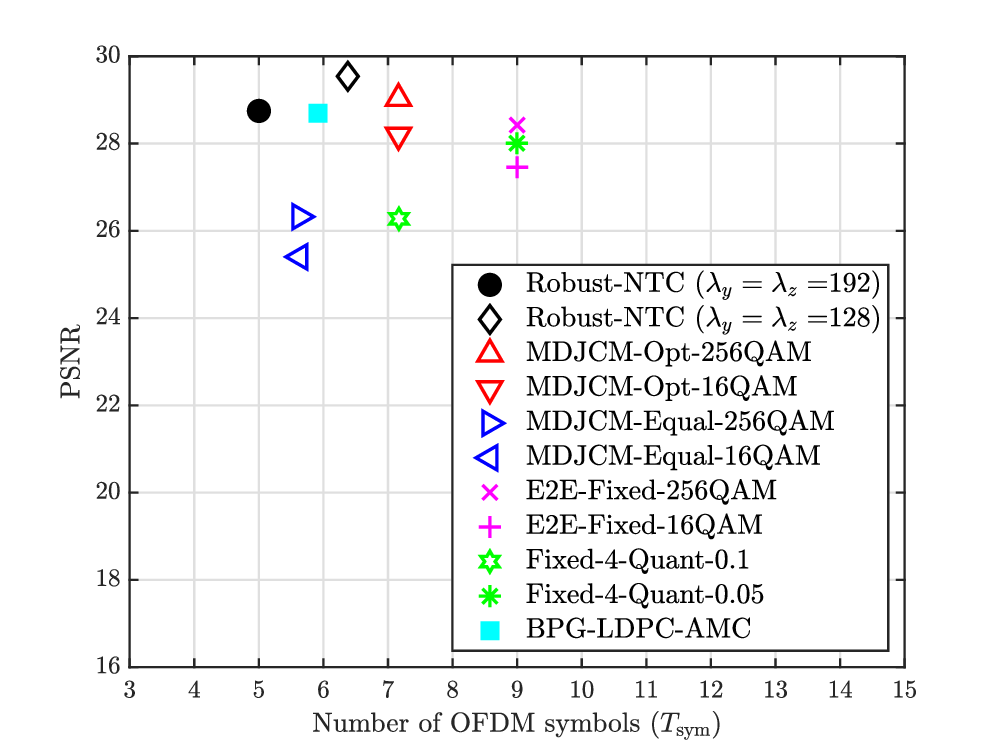, width=6.5cm}}\hspace*{-6mm}
        \subfigure[$\mathrm{SNR}=15$ dB]
    {\epsfig{file=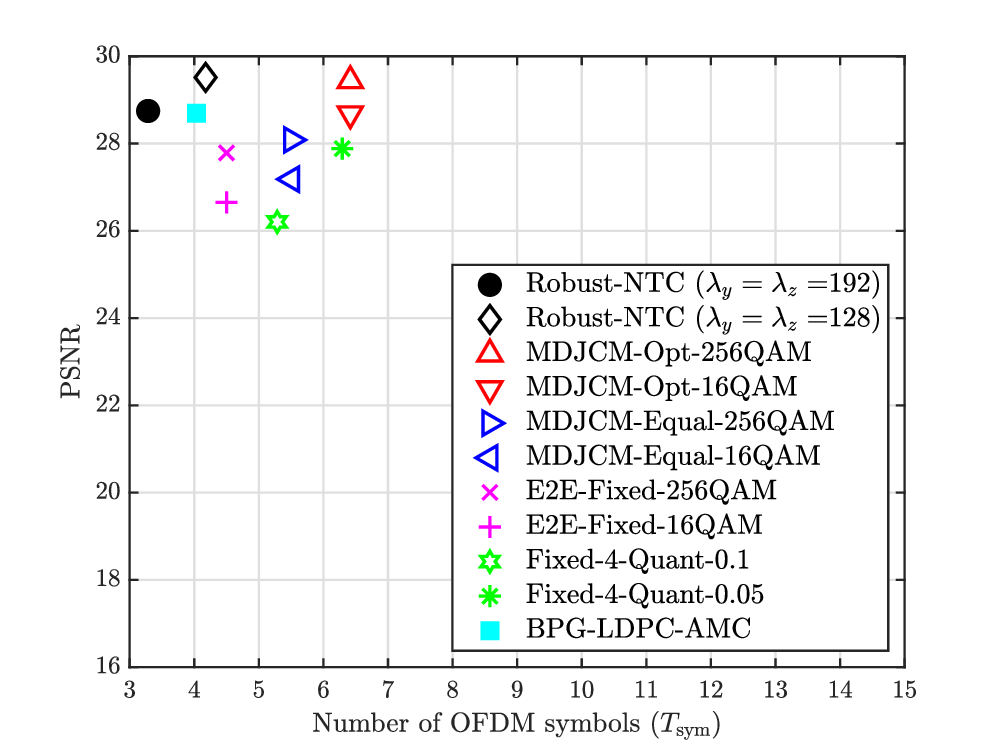, width=6.5cm}}\hspace*{-6mm}
        \caption{\textcolor{black}{Comparison of the proposed Robust-NTC and baseline schemes in terms of PSNR versus the number of OFDM symbols under different SNR levels, evaluated on the STL-10 dataset.}}
        \label{fig:diff_SNR_PSNR_STL}}
    \end{minipage}
\end{figure*}

\item\textcolor{black}{\textbf{BPG-LDPC-AMC~\cite{LDPC_AMC_table}:}
This baseline represents a standard separated coding pipeline. It compresses the source image using the BPG codec and encodes the resulting bitstream using 3GPP 5G NR LDPC codes with a fixed codeword length of $n_c=4096$. The adaptive modulation and coding (AMC) configuration is selected from 3GPP TS 38.214, Table 5.1.3.1-1~\cite{3gpp_table}.
For frequency-selective OFDM channels, AMC is performed based on the exponential effective SNR mapping (EESM)~\cite{LDPC_AMC_table}. The effective SNR $\gamma_{\mathrm{eff},t}$ for the $t$-th OFDM symbol is computed as
$$
\gamma_{\mathrm{eff},t}
= -\beta \ln \!\left( \frac{1}{N_{\mathrm{sc}}}
\sum_{j=1}^{N_{\mathrm{sc}}}
\exp\!\left(-\frac{\bar \gamma_{t,j}}{\beta}\right) \right),
$$
where $\bar \gamma_{t,j}$ denotes the received SNR of the $j$-th subcarrier in the $t$-th OFDM symbol under equal power allocation across subcarriers, and $\beta$ is the EESM calibration parameter associated with each modulation and coding scheme (MCS), with the corresponding values taken from Table II in~\cite{LDPC_AMC_table}.
To satisfy the target end-to-end block error rate (BLER) $P_{\text{target}}=0.001$ across $N_c$ segmented LDPC blocks, the AMC module dynamically selects the highest MCS index whose expected BLER satisfies the adjusted per-block target\cite{3gpp_table, LDPC_AMC_table}:
$$
\mathrm{BLER}_{\mathrm{target}} = 1 - (1 - P_{\text{target}})^{1/N_c}.
$$
At the receiver, decoding is performed using a log-likelihood-ratio (LLR)-based belief propagation (BP) algorithm with 20 iterations. If BPG decoding fails, the entire image is reconstructed using the median pixel value prior to measuring the distortion.}
\end{itemize}
For fair comparison, all baseline models are configured with the same Swin-Transformer backbone with a larger model size.
Except for {E2E-Fixed-$M$QAM}, which adopts a fixed OFDM symbol length, all other schemes operate with variable-length transmission.
To enable effective performance comparison, each channel realization is configured to transmit $16$ test images, and the rate cost per image is computed as the average number of OFDM symbols required per transmission.
Unlike the Robust-NTC, MDJCM-$M$QAM requires transmission of the rate tensor as essential side information, while $\mathbf{\hat z}$ is optionally transmitted depending on the target rate.
To ensure fairness, only the rate-token information is transmitted in the MDJCM-$M$QAM baselines, and its symbol overhead is compensated according to the capacity estimated under the equal-SNR constraint, as in {MDJCM-Opt-$M$QAM}, which determines the optimal number of active subcarriers.
The same compensation is applied to the Robust-NTC to maintain consistent accounting of side-information transmission overhead.
\textcolor{black}{Table~\ref{tab:side_info_bits} reports the average number of side-information bits per image. Robust-NTC requires fewer or comparable side-information bits than the MDJCM-$M$QAM baseline across the considered datasets, owing to entropy modeling of side information.}
\textcolor{black}{Furthermore, the side-information overhead grows moderately relative to the increase in image scale, demonstrating the scalability of the proposed framework.}

\begin{figure*}[t]
    \centering 
    \begin{minipage}{2\columnwidth}
        \centering 
        {\setlength{\fboxrule}{0pt}
        \subfigure[$\mathrm{SNR}=5$ dB]
         {\epsfig{file=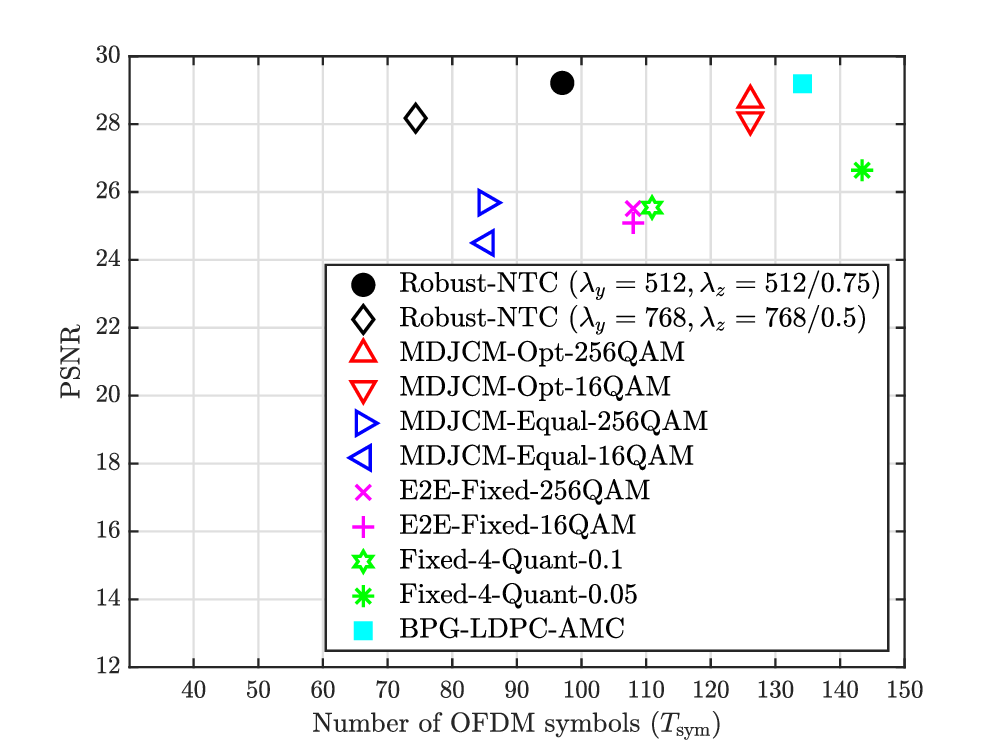, width=6.5cm}}\hspace*{-6mm}
        \subfigure[$\mathrm{SNR}=10$ dB]
    {\epsfig{file=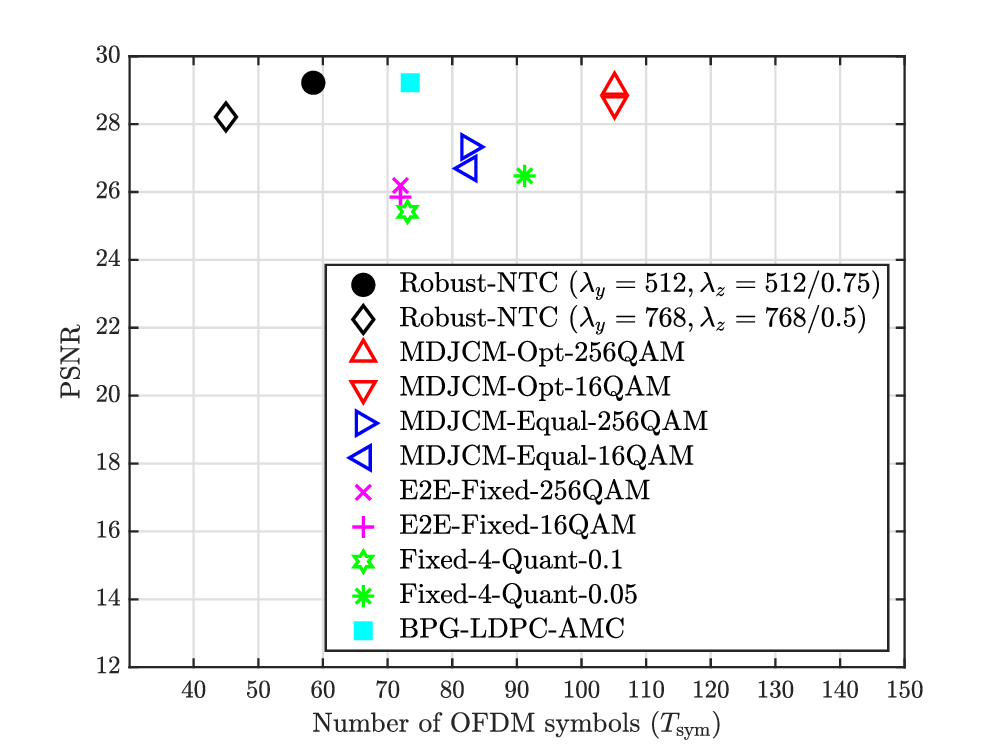, width=6.5cm}}\hspace*{-6mm}
        \subfigure[$\mathrm{SNR}=15$ dB]
    {\epsfig{file=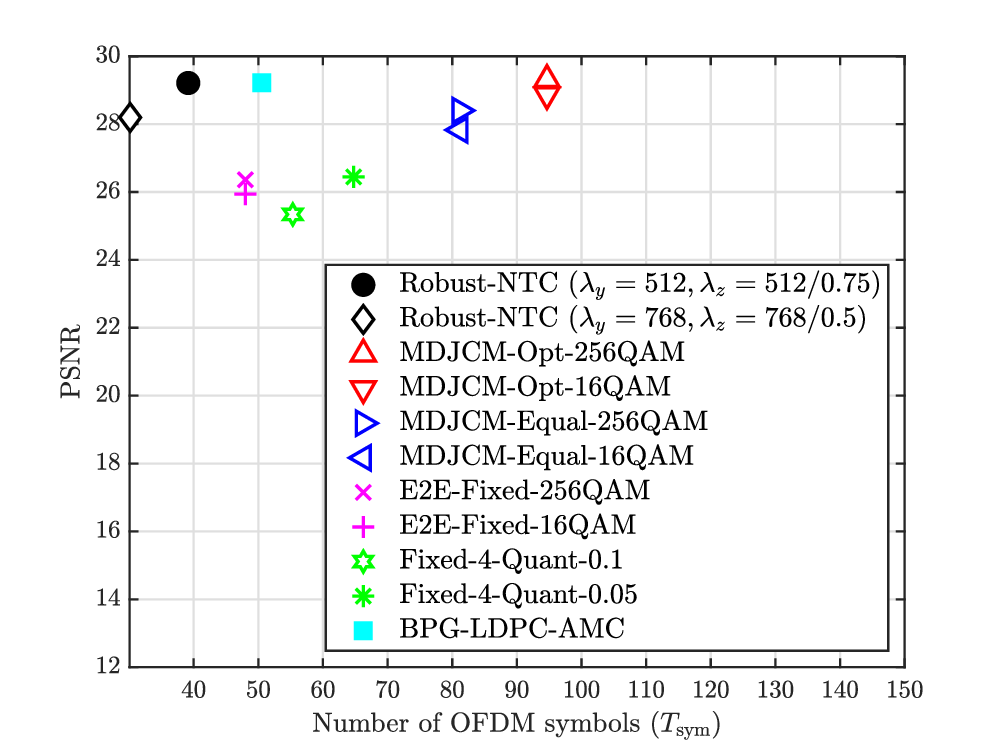, width=6.5cm}}\hspace*{-6mm}
        \caption{\textcolor{black}{Comparison of the proposed Robust-NTC and baseline schemes in terms of PSNR versus the number of OFDM symbols under different SNR levels, evaluated on the Kodak dataset.}}
        \label{fig:diff_SNR_PSNR_Kodak}}
    \end{minipage}
\end{figure*}

\begin{figure}[t]
    \centering
    \includegraphics[width=1\columnwidth]{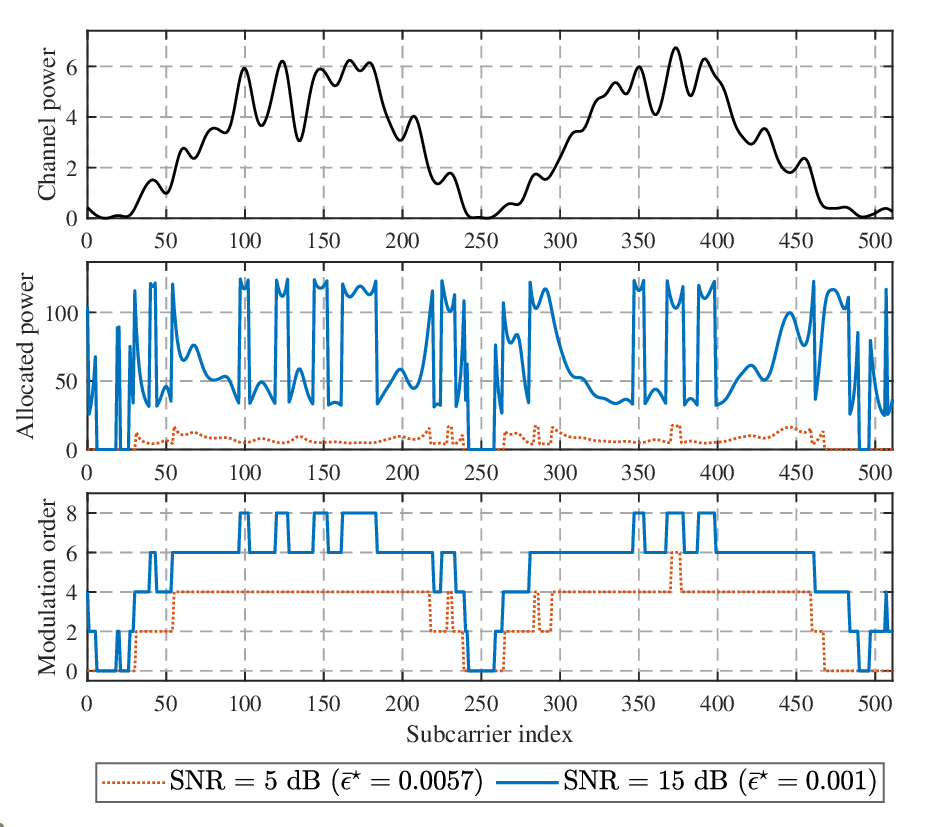}
    \caption{Subcarrier-wise adaptive bit and power allocation of the Robust-NTC $(\lambda_y=\lambda_z=128)$ under frequency-selective fading, evaluated on the CIFAR-10 dataset.
    The figure shows the channel power (top), allocated power (middle), and modulation order (bottom) for SNR levels of 5~dB and 15~dB.}
    \label{fig:Subcarrier_AMC_comp}
\end{figure}

\begin{figure}[t]
    \centering
    \includegraphics[width=1\columnwidth]{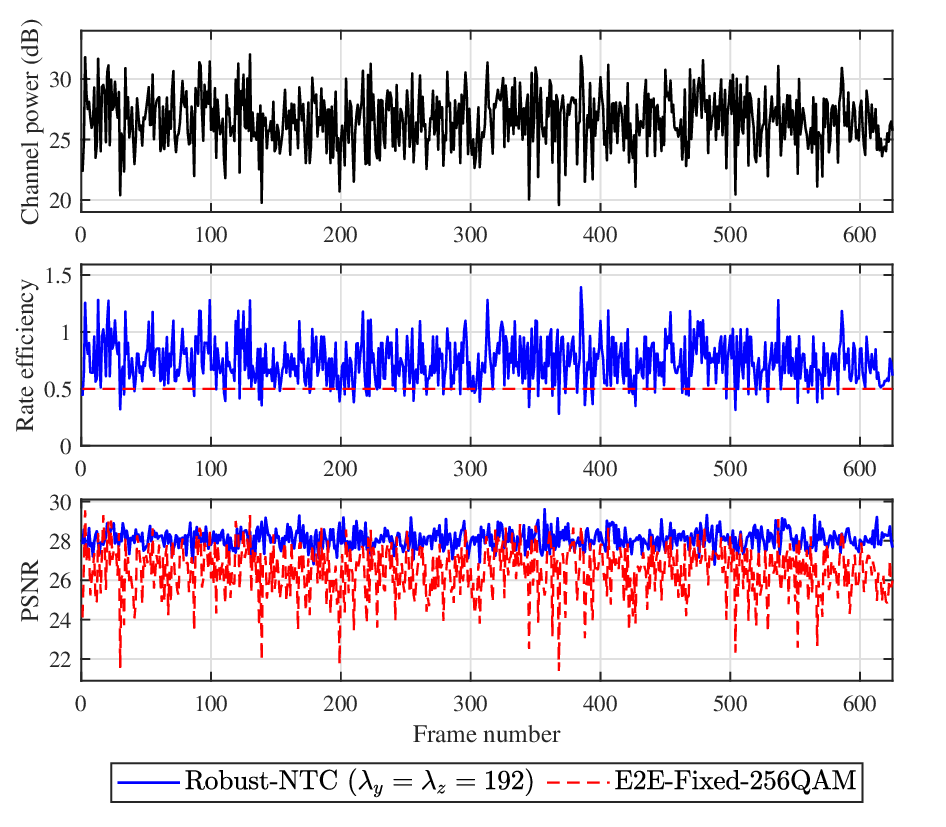}
    \caption{Frame-wise comparison of total channel power, rate efficiency, and PSNR under 5~dB SNR, evaluated on the CIFAR-10 dataset using Robust-NTC $(\lambda_y=\lambda_z=192)$ and E2E-Fixed-$256$QAM.
    The figure shows the total subcarrier channel power in dB (top), the rate efficiency defined as $1/T_{\mathrm{sym}}$ (middle), and the resulting PSNR (bottom) across transmitted frames.}
    \label{fig:5dB_RDFrame}
\end{figure}

\begin{figure*}[t]
    \centering 
    \begin{minipage}{2\columnwidth}
        \centering 
        \hspace*{-5.5mm}
        {\setlength{\fboxrule}{0pt}
        \subfigure[$\mathrm{SNR}=5$~dB]
         {\epsfig{file=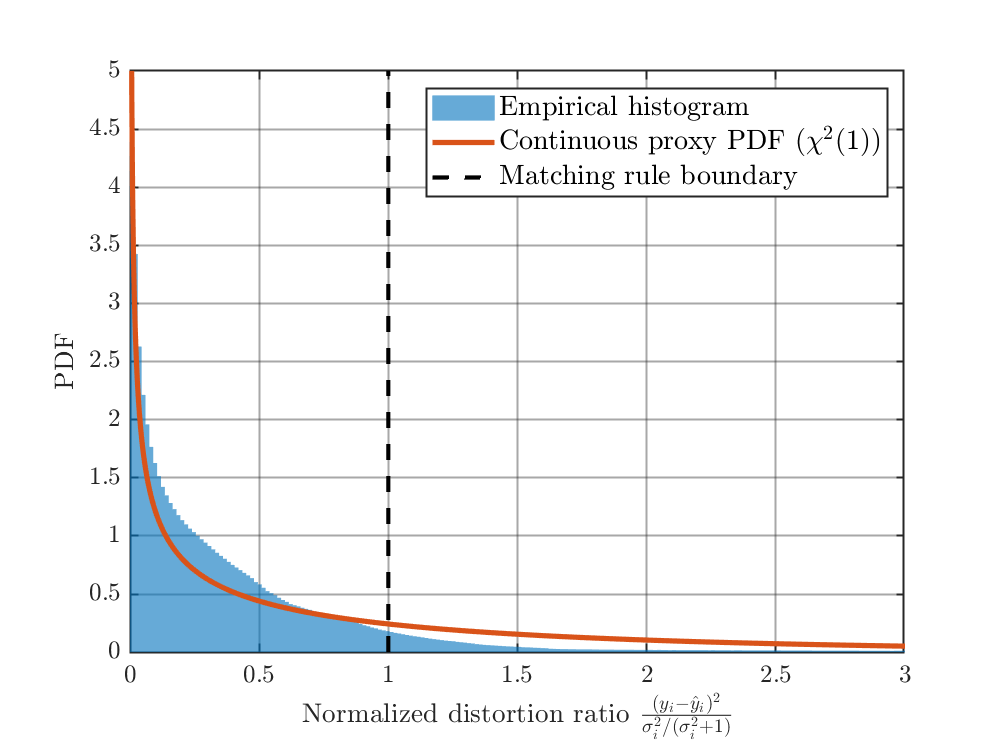, width=6.5cm}}\hspace*{-6mm}
        \subfigure[$\mathrm{SNR}=10$~dB]
         {\epsfig{file=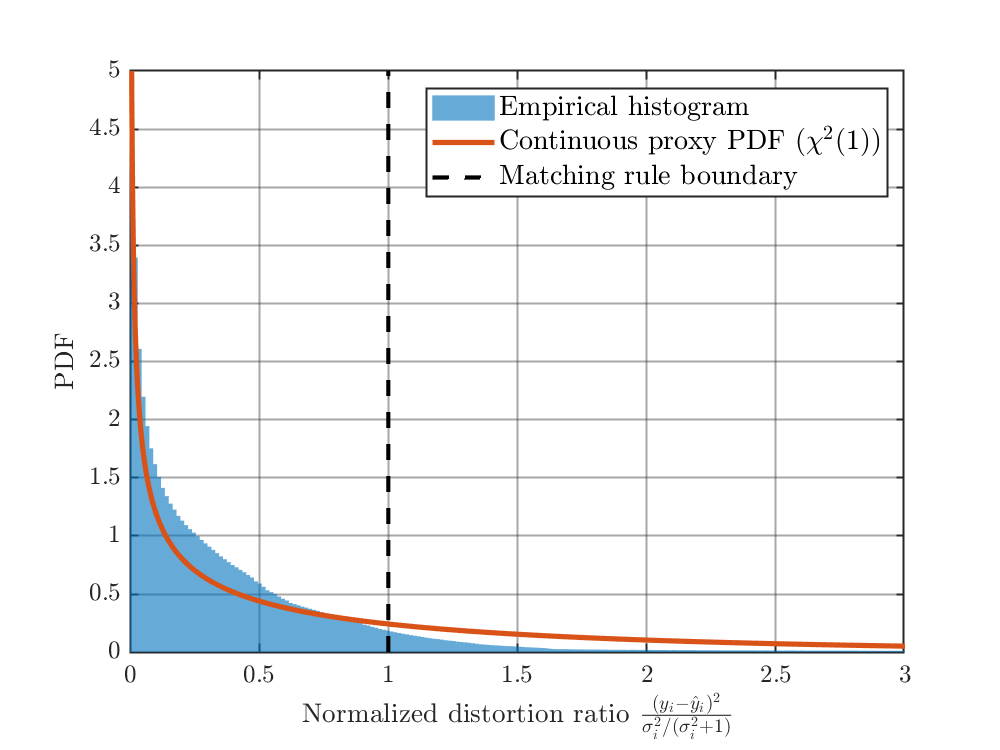, width=6.5cm}}\hspace*{-6mm}
        \subfigure[$\mathrm{SNR}=15$~dB]
         {\epsfig{file=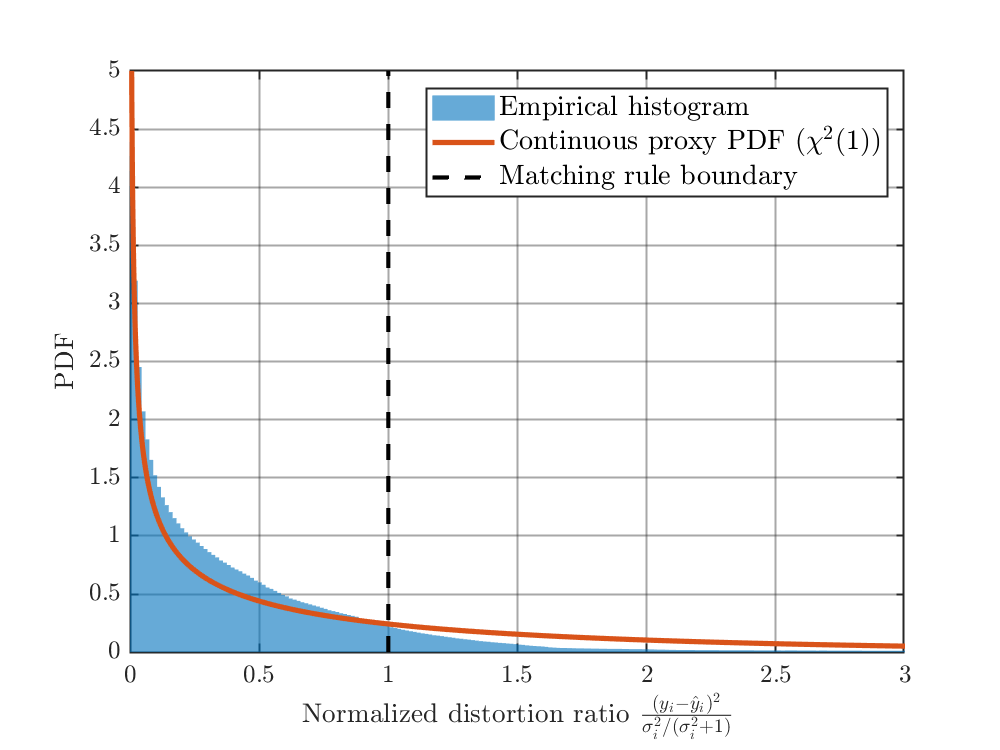, width=6.5cm}}\hspace*{-6mm}
        }
        \caption{\textcolor{black}{Empirical PDFs of the normalized distortion ratios $(y_i - \hat{y}_i)^2 / \big(\sigma_i^2 / (\sigma_i^2 + 1)\big)$ compared against the continuous proxy reference $\chi^2(1)$ under different SNR levels, evaluated on the STL-10 dataset in the proposed Robust-NTC framework ($\lambda_y=\lambda_z=128$).}}
        \label{fig:elementwise_pdf}
    \end{minipage}
\end{figure*} 

The proposed Robust-NTC framework models the latent variables to follow a Gaussian distribution conditioned on the learned hyperprior parameters.
Fig.~\ref{fig:Latent_CDF} presents the empirical CDFs of the normalized latents $(y_i-\mu_i)/\sigma_i$ obtained from the CIFAR-10 test dataset. 
To ensure that the analysis reflects statistically meaningful latent components, only elements satisfying $\sigma_i \ge \sigma_{\min}$ are considered, where $\sigma_{\min}$ serves as a lower variance threshold.
As shown in the figure, the empirical CDFs exhibit a close alignment with the Gaussian reference, confirming that the learned hyperprior effectively regularizes the latent representation toward Gaussianity. 
Moreover, increasing $\sigma_{\min}$ results in an even tighter match, indicating that high-variance latent dimensions follow the Gaussian prior more closely. 
This observation validates the Gaussian-modeling assumption employed in the channel-aware quantizer design and demonstrates that the proposed hyperprior formulation yields statistically consistent latent variables suitable for robust transmission over noisy conditions.

Fig.~\ref{fig:diff_SNR_PSNR} compares Robust-NTC with baseline schemes on CIFAR-10 in terms of PSNR versus the number of OFDM symbols across different SNR levels.
Overall, the proposed method achieves more efficient rate-distortion operating points than all baselines, maintaining comparable PSNR with reduced OFDM symbol usage.
With distortion matching and hyperprior-based adaptation, Robust-NTC maintains nearly uniform PSNR while substantially reducing OFDM symbol usage as the SNR increases.
Fixed-modulation baselines show improved PSNR with larger modulation orders, with the gap widening at higher SNRs.
Among the MDJCM-based baselines, the MDJCM-Equal-$M$QAM provides reasonable rate efficiency but suffers PSNR degradation at low SNR, since enforcing equal effective channel gains across subcarriers lowers overall power efficiency.
In contrast, MDJCM-Opt-$M$QAM enhances robustness through adaptive subcarrier selection, at the expense of higher rate overhead.
The E2E-Fixed-$M$QAM model employs a fixed symbol length and modulation order, which limits adaptability to varying channels. Despite being trained under the corresponding channel and SNR setting, it shows inferior performance compared with Robust-NTC.
Although the Fixed-$b$-Quant-$\epsilon$ schemes also employ Algorithm~1 to adapt their rate to the channel and SNR conditions, their uniform quantizers with fixed $\epsilon$ disregard latent-channel statistics, resulting in limited rate efficiency and weaker robustness compared with Robust-NTC.
For instance, Robust-NTC with $(\lambda_y=\lambda_z=192)$ achieves PSNR performance comparable to that of Fixed-$4$-Quant-$0.05$ while requiring less than half the number of OFDM symbols, demonstrating a significant rate advantage under comparable reconstruction quality.
\textcolor{black}{The BPG-LDPC-AMC baseline shows stable average PSNR while requiring fewer OFDM symbols as the SNR increases, due to fine-grained MCS adaptation and the strong decoding capability of long LDPC blocks. Nevertheless, it exhibits a less favorable rate-distortion trade-off than Robust-NTC, especially at low and moderate SNRs.}

Fig.~\ref{fig:diff_SNR_PSNR_STL} shows the PSNR performance of the proposed Robust-NTC and baseline schemes on the STL-10 dataset.
Compared with CIFAR-10, the larger image size and higher content variability of STL-10 make the latent representation more heterogeneous and challenging to compress efficiently.
Nevertheless, Robust-NTC maintains a stable rate-distortion tradeoff across all SNR levels, demonstrating that its learned latent modeling and distortion-matching strategy generalize well to higher-resolution data.
Consistent with the CIFAR-10 results, the proposed framework preserves a similar performance trend relative to the baselines, achieving comparable reconstruction quality with substantially fewer OFDM symbols.
This result highlights that Robust-NTC effectively scales to more complex image structures, maintaining high rate-distortion efficiency and adaptive transmission behavior under increased source dimensionality.

\textcolor{black}{Fig.~\ref{fig:diff_SNR_PSNR_Kodak} illustrates the performance of the proposed framework on the Kodak dataset. Although the Kodak images are $384$ times larger than those in CIFAR-10, and the evaluation relies on a cross-dataset setup (trained on ImageNet validation dataset patches, tested on full-resolution Kodak images), Robust-NTC continues to exhibit highly competitive rate-distortion efficiency across all evaluated SNRs. This suggests that the proposed distortion-matching-based, latent-channel-aware adaptation strategy generalizes well to unseen, high-resolution data. In contrast, the limitations of the baseline schemes---previously noted on the smaller datasets---are more clearly amplified here. The MDJCM-based methods suffer performance degradation, as their sub-optimal adaptation to frequency-selective OFDM channels struggles to manage large, diverse data blocks efficiently. The other baselines, including BPG-LDPC-AMC, also remain inferior to Robust-NTC in overall rate-distortion efficiency.}

Fig.~\ref{fig:Subcarrier_AMC_comp} illustrates the subcarrier-wise adaptive bit and power allocation performed by the proposed Robust-NTC framework at SNR levels of 5~dB and 15~dB.
In general, subcarriers with higher channel power are assigned larger transmit power and higher modulation orders, while weaker subcarriers receive smaller power and lower-order modulations.
This allocation behavior follows the theoretical water-filling trend \cite{Waterfilling}, while the power is further optimized to equalize the effective SNR among subcarriers sharing the same modulation order, resulting in an inverse power allocation pattern that enhances practical stability.
Note that the optimal BER targets $\bar{\epsilon}^\star$, determined by Algorithm~3, are $0.0057$ and $0.001$ for 5~dB and 15~dB, respectively.
Under favorable channel and total power conditions (e.g., 15~dB), the system assigns lower $\bar{\epsilon}^\star$ and higher modulation orders, thereby improving rate efficiency.
Conversely, in adverse channel conditions (e.g., 5~dB), higher $\bar{\epsilon}^\star$ and lower modulation orders are allocated to enhance robustness, demonstrating the effectiveness of the proposed adaptive transmission strategy.

Fig.~\ref{fig:5dB_RDFrame} compares the rate efficiency and PSNR of Robust-NTC $(\lambda_y=\lambda_z=192)$ and E2E-Fixed-256QAM under 5~dB SNR as the total channel power varies across frames.
Here, the rate efficiency is defined as $1/T_{\mathrm{sym}}$, representing the inverse of the number of OFDM symbols required for transmitting one image frame.
As the total channel power fluctuates across frames, Robust-NTC adaptively adjusts the transmission rate by controlling $T_{\mathrm{sym}}$, thereby maintaining consistently higher rate efficiency and more stable PSNR performance.
In contrast, E2E-Fixed-$256$QAM employs a fixed symbol length and modulation order, leading to significant PSNR fluctuations as the channel condition varies and exhibiting limited robustness.
These results highlight the effectiveness of the proposed adaptive rate control in achieving both transmission efficiency and robustness under dynamic channel conditions.

\textcolor{black}{
Fig.~\ref{fig:elementwise_pdf} compares the empirical PDFs of the normalized distortion ratios across different SNR levels against the continuous proxy reference to verify that the proposed framework effectively accounts for the full impact of channel impairments and ensures reliable operation.
To ensure the analysis reflects meaningful features, only active latent elements satisfying $\sigma_i^2 > \delta$ are included in the evaluation. 
Under the Gaussian proxy, the normalized squared error for the $i$-th element is formulated as a Chi-squared distribution with one degree of freedom, given by
\begin{equation}
    \frac{(y_i - \hat{y}_i)^2}{\sigma_i^2 / (\sigma_i^2 + 1)} \sim \chi^2(1).
\end{equation}
As shown in the figure, the empirical distributions across all SNR regimes are broadly consistent with the theoretical $\chi^2(1)$ curve, while consistently exhibiting a lighter tail. 
Notably, severe distortion events near or beyond the matching-rule boundary of $1$ occur less frequently than predicted by the continuous proxy model. 
This observation confirms that the adopted Gaussian proxy provides a conservative approximation of the actual discrete distortion induced by quantization and channel errors. 
Consequently, the proposed distortion-matching rule effectively allocates resources with a practical safety margin, improving robustness against severe latent distortion under practical channel fading.}

    \section{Conclusion}\label{sec:Conclusion}
This paper presented a generalizable digital JSCC framework that combines variational latent modeling with optimization-driven adaptive transmission. 
By introducing a Gaussian-proxy formulation within a hyperprior-based VAE architecture, the method explicitly learns element-wise latent statistics that capture heterogeneity in uncertainty and distortion sensitivity, without assuming a specific channel during training.
A distortion-matching mechanism then optimizes quantization and bit depth to preserve learned distortion targets. 
For practical deployment, Robust-NTC was integrated into an OFDM system with cross-layer optimization of quantization, modulation, power, and OFDM symbol allocation. 
Experiments over practical frequency-selective fading channels show that the proposed method maintains stable reconstruction quality while substantially reducing OFDM symbol usage across a wide range of SNRs.

\textcolor{black}{Several directions remain for future work. As the current framework adopts practical design simplifications, an important extension is to develop more refined designs that better approach the optimal joint solution while preserving computational efficiency. One promising direction is to explore more suitable source and proxy distribution models that better capture the natural statistics of JSCC latents, thereby further improving rate-distortion efficiency. Another direction is to develop more efficient channel allocation strategies beyond a common target-BER design, for example by grouping latent elements with similar variances, as well as more effective quantizer designs based on unequal error protection within the assigned bit budget. It is also of interest to move beyond the BSC-based abstraction and design quantizers directly from soft LLR statistics \cite{soft_quant}, and to replace the current greedy procedure with more computationally efficient resource-allocation methods. Finally, extending the proposed framework to broader settings, such as MIMO and multi-user systems, would be a valuable direction for future work.}

\appendices

\bibliographystyle{IEEEtran}
\bibliography{Reference}
    
\end{document}